\documentclass[]{article}

\usepackage{alanstyle}
\usepackage{graphicx}
\usepackage{empheq}
\usepackage{apacite}
\usepackage{tikz-cd}
\title{Option-based Equity Risk Premiums}
\author{Alan L. Lewis\footnote{Newport Beach, California, USA; email: alewis@financepress.com}}

\begin{document}

\maketitle

\begin{abstract}
	We construct the term structure of the (forward-looking, US market) equity risk premium from SPX option chains.
	The method is ``model-light". Risk-neutral probability densities are estimated by fitting $N$-component Gaussian 	mixture models to option quotes, where $N$ is a small integer (here 4 or 5). These densities are transformed to their real-world equivalents by exponential tilting with a single parameter: the Coefficient of Relative Risk Aversion $\kappa$. From history, I estimate $\kappa = 3 \pm 0.5$. From the inferred real-world densities, the equity risk premium is readily calculated. Three term structures serve as examples.

\end{abstract}

\section{Introduction and summary}

The equity risk premium (ERP throughout) has frequently been called the most important number
in finance. To be fair, so have LIBOR and the 10-year US Treasury yield.  It's certainly the elusive one: interest rates are readily observed, while the ERP (as defined here) is a forward-looking 
``market" expectation, requiring estimation. 

US market history suggests a long-run (unconditional) ERP of 4-6\% per year.
Our topic is the difficult problem of estimating time variation and term structure effects. Plausibly,
key drivers of this \emph{conditional} ERP are outlooks for inflation, rates and yields, earnings, and volatility.

There are many models and approaches in the literature. An interesting and diverse collection of views
is found in \cite{rethink:2011}.  In \cite{duarte:2015}, the authors use the best linear
combination of 20 models to produce, among other things, ERP term structure estimates at various dates.  
In agreement with them, we find that the ERP term structure has significant time variation and can be flat, upward, or downward
sloping. Later, in Sec. \ref{sec:ERPresults}, we further compare our results to theirs.

Many authors use time series models to infer an ERP from the (weak) predictive power of financial ratios: dividend yields or PE ratios. However, there is controversy over the predictive power. Even without the controversy, a difficulty is that financial ratios don't vary much in the short-term. Another difficulty: beyond key drivers is the large and nebulous set $I_t$, the time $t$ ``state-of-the-world" -- any element of which may become (momentarily) important to the market's outlook. For example, at this writing, trade negotiations have become a market focus.\footnote{\label{ft:COVID}Update: at this revision, the COVID-19 pandemic is driving everything. Using the methods here, the US ERP during the pandemic is studied in \cite{lewis:2020a}.}   

Here, we develop a new approach, using SPX options and a one-parameter change-of-measure to estimate 
the ERP term structure at arbitrary trade dates. 
It's relatively objective, driven largely by forward-looking market data: option quotes. 

While certainly not ``model-free", I  call it ``model-light". No stochastic process is adopted for SPX. No financial ratios are singled out as explicit drivers of the ERP, although they can certainly play an implicit role in market expectations. Indeed, \emph{anything} the market deems momentarily important to its
outlook, as long as it's reflected in the options market, can influence our ERP estimates. Our key modelling assumption --
which has theoretical and historical support -- is that an exponential change-of-measure transforms the option-based risk-neutral density (RND) to the real-world density (RWD).   

\begin{figure}[hp] 
  	\caption{\bf{The CBOE's VIX index: 2018-2019}}
  	\vspace{10pt}
  	\begin{center}
  		\includegraphics[width=0.8\textwidth]{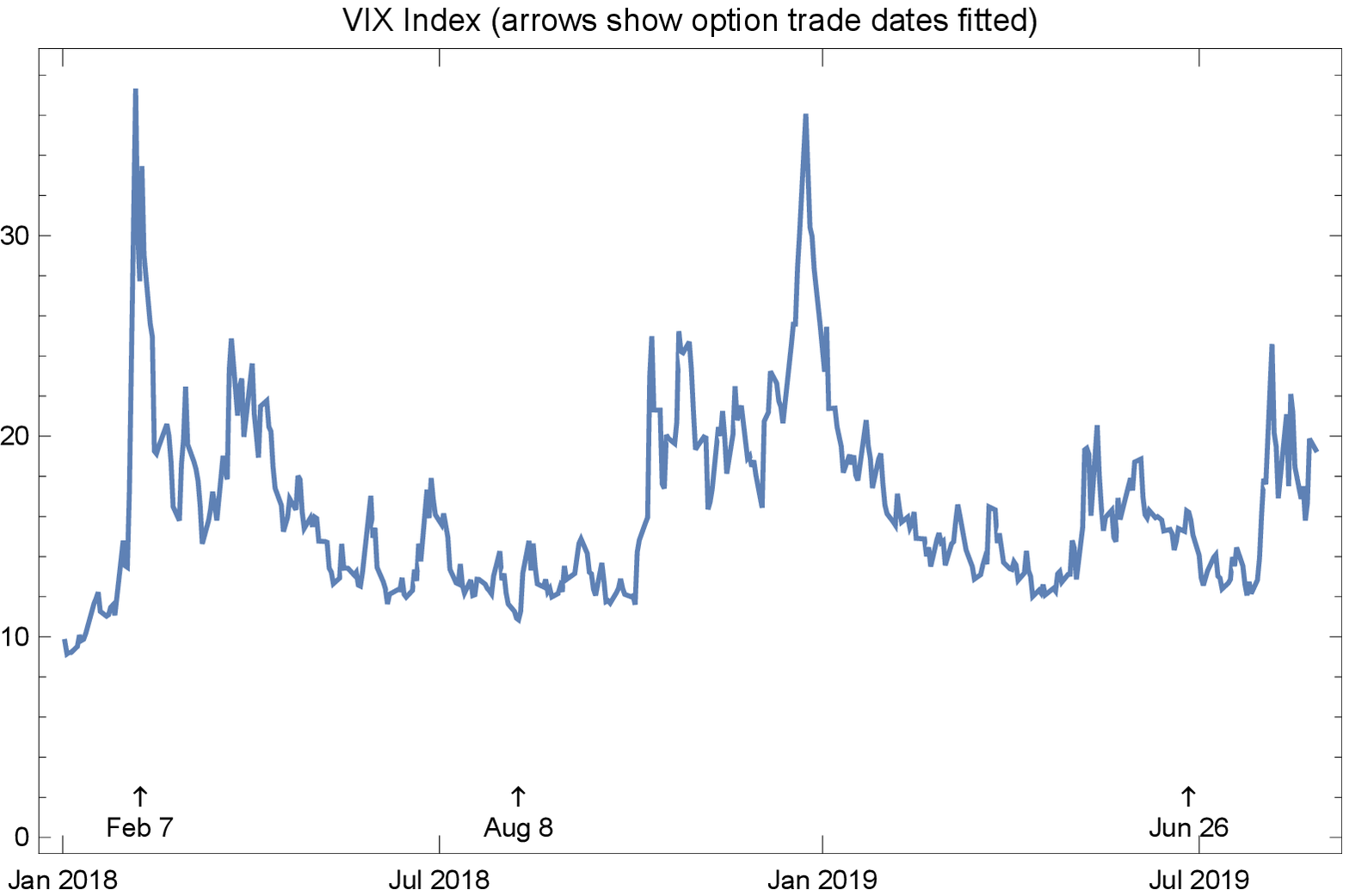}
  		\newline\newline
  	\end{center}		
  	\label{fig:VIXplot}
  \end{figure}

Figs. \ref{fig:ERPplot1}-\ref{fig:ERPplot3} show three examples of our new ERP term structure estimates. Near expirations
are closely spaced in time, and made more visible with log time scales. The trade dates shown have relatively high, low, and middle-of-the-road (recent) volatility environments, as measured by the VIX index: see Fig.\ref{fig:VIXplot}. In the higher volatility environment, the ERP estimates decay
 from a large value $\approx 26\%$ (from 2-day options) toward the longer-run averages (from 2.5 year options). 
 Conversely, in the lower volatility environment, the ERP term structure is increasing with the term, again toward unconditional values. Finally, the last volatility environment yields a relatively flat term structure. 
 Qualitatively, those are plausible term structure behaviors, key drivers here being volatility and risk-aversion. The decay from 26\% is just a recent example. More generally, volatility and the short-term (annualized) ERP estimates can be expected to be very high in systematically stressed markets -- let's say with
 VIX above 40 -- as was seen in the Financial Crisis of 2007-2008.
 
 The figure ERP bands use a Coefficient of Relative Risk Aversion, $\kappa$, in the range $2.5 - 3.5$,
 with  $\kappa=3.0$ being the central estimate (central curve). 
  We argue for that range in Sec. \ref{sec:kappaestimates}, using market history. Note that $\kappa$ also serves as 
  the parameter for the exponential change-of-measure. Theoretically, the exponential form follows from a 
 standard equilibrium market-clearing argument using a representative investor with power utility  -- an argument 
 reviewed in the Appendix. To estimate $\kappa$, we use the fact that it is the exponential tilt parameter that turns a (long-run historical) 
excess return sequence into a zero-mean sequence: see eqn (\ref{eq:kappaestimator}) below. Looking forward with options data (in Sec. \ref{sec:OIERP}), we estimate RND's by fitting SPX option quotes to Gaussian mixture models. Then, $\kappa$ effects a
one-parameter change of measure from the RND's to RWD's, the real-world probability densities.
Examples are shown in Figs. \ref{fig:pdfsA}-\ref{fig:pdfsC}. Given the RWD's, the
ERP's are readily calculated. That's the paper in a nutshell.

 For the remainder of this section, I give a more
elaborated summary. Key formulas are boxed.

\subsection{Some definitions and notations}
What exactly do we mean by the ERP? There are many closely related definitions.  First of all, we  adopt a US perspective, so ``equities" means a broad-based (capitalization weighted) measure of the entire US equity market, frequently proxied by the S\&P 500 Index. Second, the ERP is far from a single number. Like interest rates, it's time-varying, with a time-varying term structure.
With $\EBB_t$ denoting a (real-world) expectation conditional on  date-$t$ information $I_t$ -- broadly speaking: the ``state of the world" -- we define:

\be   \mbox{ERP}_{t,T}  = \Et{R^e_{t,T}} - R^f_{t,T} = \Et{R^e_{t,T} - R^f_{t,T}}, \quad \mbox{where at time $t$}: \label{eq:ERPdef}  \eb
\begin{itemize}
   \item  $R^e_{t,T}$ is a future random \emph{total} return on the equity market from $t$ to $T$, and
   \item  $R^f_{t,T}$ is a time-$t$ observable risk-free return (using US Treasury instruments).
\end{itemize}
Returns in (\ref{eq:ERPdef}) are simple total returns: $R^e_{t,T} = (\bar{S}_T - \bar{S}_t)/\bar{S}_t$, where $\bar{S}$ is a total-return index incorporating reinvested dividends. (Without a bar, $S_t$ is the price series without dividends). Call $R^e_{t,T} - R^f_{t,T}$ the \emph{excess total return}. 
Like interest rates, we'll always give estimated ERP's on an \emph{annualized percentage basis}. For those, we multiply the ERP calculated from (\ref{eq:ERPdef}) by $100 \times f_{ann}$, where the annualization factor $f_{ann} = 1/(T-t)$, with time measured in years.  
That convention is used in Figs. \ref{fig:ERPplot1}-\ref{fig:ERPplot3} and associated tables.
 
Let's call $R^f_{t,T}$ the \emph{risk-free basis} for the ERP. There are two natural choices in the literature for the risk-free basis.  One choice ignores the $T$ and use a short-maturity US Treasury bill return for each basis. The second choice, which we employ,  uses the time-$t$ return available from Treasury instruments that mature at $T$. That's more natural for our forward-looking, option-based estimates, and is used in Figs. \ref{fig:ERPplot1}-\ref{fig:ERPplot3}. 
 For our history-based estimates of $\kappa$, $R^f_{t,T}$ is a realized return. Those are constructed from 
 monthly Tbill returns (when $T-t$ is one month), chaining together monthly returns (when $T-t$ is greater than a month), or pro-rating monthly returns (when $T-t$ is one day).

 \begin{table}[t] 
	\caption{{\bf{Some methods for estimating the RND}}}
	\begin{center}
		\begin{tabular}{ll} 			
		    Reference    & Description   \\
		    	\cmidrule(r){1-1}   \cmidrule(r){2-2} 
		    \cite{bliss:2004}   &  Smoothing spline fit to mid-quote IV's $+$ flat (Gaussian) extensions   \\
		    \cite{gatheral:2013} &  5-parameter SVI fit to mid-quote IV's implies a full RND                 \\
		    \cite{fengler:2005}  &  Smoothing spline fit to option prices implies a partial RND               \\
		    \cite{figlewski:2010} & Weighted spline fit to mid-quote IV's + GEV distribution extensions               \\		  
		    \cite{malz:2014}  &   Interpolating spline fit to mid-quote IV's $+$ flat (Gaussian) extensions            \\
			\bottomrule 
		\end{tabular}
	\end{center}
	\label{tab:RNDmethods}
\end{table}

\subsection{Estimating risk-neutral and real-world probability densities}

 Let $Q_{t,T}(S_T)$, denote the time-$t$ RND for $S_T$, where $S_T$ is the terminal index price.
 The RND depends upon $S_t$, the starting  index price, and generally $I_t$, our catch-all
 for information known at $t$. These latter dependencies are freely suppressed, but implied by the $t$ subscript
 in  $Q_{t,T}(S_T)$.
 
From Treasury rates, infer deterministic risk-free discount factors 
$D_{t,T} \equiv  1/(1 + R^f_{t,T}) = \exp\{-r_{t,T} (T-t)\}$, using various of our notations. Then, where $K$ is the option strike price, call option values $C_{t,T}(K)$ are  given by   
\be C_{t,T}(K)= D_{t,T} \, \Eqt{(S_T - K)^+} = D_{t,T} \int_0^{\infty} (S_T - K)^+ Q_{t,T}(S_T) \, dS_T,
\label{eq:cval} \eb 
where $\Eqt{\cdots} \equiv \EtwoQ{\cdots}{I_t}$. 
 Given a set of option prices, $\{C_{t,T}(K_i): i=1..N_{opts}\}$, there are various \underline{types} of approaches to estimating  $Q_{t,T}(S_T)$:
  \begin{itemize}
  	\item[(1)] Modelling the underlying stock process.
  	\item[(2)] Fitting approaches suggested by the Breeden-Litzenberger relation.
  	\item[(3)] Direct modelling of the $\QBB$-density $Q_{t,T}(S_T)$ or (as here) the pdf for $\log S_T$.  
  \end{itemize}
  \pbold{Type 1.} A typical framework is the following. Postulate a (continuous-time) $\QBB$-measure jump-diffusion process,
  $dS_t/S_t = (r_t - \delta_t) \, dt + \sigma_t \, dW_t + dJ_t$, where $\sigma_t$ is a parametrized
  stochastic volatility process and $J_t$ is a parametrized (zero-mean) jump process.
  Free parameters are estimated by fits to option prices, implied volatilities, or similar
  targets. Parameters estimates yield $Q_{t,T}(S_T)$'s. 
      
  The main problem is that stationary models will have, let's say,  6-12 parameters for
  computational tractability. Unfortunately, such a small number will prove wholly inadequate for fitting a large option data set with
  multiple trade dates. One could adopt (large) parameter sets that vary with the
  trade date $t$ and option expiration $T$, but this does not typically result in a logically consistent stochastic process. For example,
  why should the putative underlying process even \emph{know} about the arbitrary dates of option expirations?

 \pbold{Type 2.} If calls were marketed with a continuum of strikes $K \in (0,\infty)$, and zero bid-ask spreads,
then the Breeden-Litzenberger relation,
    \be Q_{t,T}(K) = \frac{1}{D(t,T)} \frac{d^2  C_{t,T}(K)}{dK^2}, \label{eq:BL} \eb 
  would yield a unique, completely ``model-free" risk-neutral density. Then, given discrete strikes and positive spreads, use interpolating or smoothing splines to fit option prices or
  implied volatilities (IV's for short). Generally, it's better to fit the $\{IV(K_i)\}$, which yields a smooth function $IV(K)$. From that, a smooth option pricing function $C_{t,T}(K) = c_{BS}(K,IV(K))$ is available, where $c_{BS}(K,\sigma)$ is the Black-Scholes formula, suppressing other arguments. Finally, (\ref{eq:BL}) and the chain rule for differentiation
  yield the estimated $Q_{t,T}(K)$. 
  
  In my opinion, the main issue with this type of approach is deciding how to extrapolate the $\{IV(K_i)\}$ to
  the ranges $0 < K < K_{min}$ and $K_{max} < K < \infty$. Here $(K_{min},K_{max})$ represent the range of
  marketable strikes. You \emph{must} extrapolate to find a proper (norm=1) RND.
  It's difficult to devise an extrapolation method that doesn't feel ad hoc. Some references for this
  type of approach are given in Table \ref{tab:RNDmethods}. 
  
  \pbold{Type 3.} Here, one parameterizes directly the RND. 
   After some experimentation with type 2 methods, I ultimately adopted a type 3 method.  
   Specifically,  with log price-returns $X_T = \log (S_T/S_t)$, the corresponding RND
   is $q_{X_T}(x) =  Q_{t,T}(S_t \e^x) S_t \e^x$.  Then, I fit a Gaussian
  Mixture Model (GMM) to option quotes.\footnote{\label{ft:thanks} My thanks to Thijs van den Berg, who suggested exploring GMM's.
  	Prior to that suggestion, I was fitting smoothing splines to IV's (similar to some
  	Table \ref{tab:RNDmethods} methods) and attempting extrapolation. But I quickly converted to
  	the GMM approach described here: it seemed a `natural' for my ERP problem setup.}  The mixture consists of $N$ Gaussians: 
  \be  \mbox{\underline{log price-returns}}: \quad q_{X_T}(x) = q_{t,T}(x) = \sum_{i=1}^N w_i \,
   \frac{\e^{-(x - \mu_i \tau)^2/(2 \sigma_i^2 \tau)}}{\sqrt{2 \pi \sigma_i^2 \tau}}, \label{eq:GMM}
       \eb
  where $\tau = T-t$, and $N$ is  a small integer (4-5 in my fits). 
  
  (Notations: strictly, I should write $X_{t,T}$ instead of $X_T$, but I don't for notational simplicity. Also,   
  I move to small letters for log-arguments).   
  
  The fitted parameters
  are $N$ positive weights, $\{w_i\}$, and  $2N$ drifts and volatilities, $\{\mu_i,\sigma_i\}$.
  After a normalization and martingale condition, this leaves $3 N - 2$ free parameters at
  \emph{each} $(t,T)$ pair associated to a trade date and an option expiration. Using (\ref{eq:cval})
  and the corresponding put value formula, free parameters are adjusted to fit option quotes -- details are found in Sec. \ref{sec:fitting}.
  
  Under an exponential change of measure, (\ref{eq:pqduality}) below, the RWD will also have Gaussian tails. At
  first glance, this might give pause because it's well-known that \emph{single} Gaussian fits to historical S\&P 500 returns are strongly rejected by statistical tests. However, we stress this rejection is irrelevant to fits of \emph{mixtures}. To see why, consider the analogous RND case. A single Gaussian implies flat IV smiles, a shape which would lead similarly
  to strong rejection of the proposed density. In contrast, a GMM can nicely fit market smiles, as we will show. 
  
  \Pbreak  
  Indeed, as I suggest in footnote \ref{ft:thanks}, in my opinion the GMM method is a `natural' for the ERP problem,
  at least as set up here. Let me summarize some of the attractive features.
  
  First,  the GMM is able to achieve good fits to option quotes.
  
  Second, by modelling a density directly, smile extrapolation is built-in and does not need to
  be grafted on -- as is the case with spline fits to IV's.

  Third, the GMM accommodates arbitrary $\kappa$ in the exponential change of measure. A plausible alternative
  might choose $q_{t,T}(x)$ from a class of models with exponential  (`semi-heavy') tails. 
  For example (although it is actually a type 2 method), J. Gatheral's SVI method has exponential tails
  \cite{gatheral:2013}. That's fine, but there 
  will be a restriction on $\kappa$ relative to the tail parameters, so that the denominators
  in (\ref{eq:pqduality}) below exist. This restriction is certainly not a deal breaker for semi-heavy tails, 
  but adds a complication if you insist on them.   
  Finally, option values and exponential changes of measure are nicely tractable under GMM.

  \pbold{Transforming to log total-return densities.}
  The cost of carry parameters are $(r_{t,T},\delta_{t,T})$. How we get them is explained later.
  Here $r_{t,T}$  is the  continuously compounded risk-free interest rate  
  and  $\delta_{t,T}$ is a continuously compounded dividend yield for the underlying stock index. Given 
  those, we have the index forward price  
  \[ F_{t,T} = S_t \exp \{(r_{t,T} - \delta_{t,T})(T-t)\}. \]  
  If the dividends were reinvested into the index, the index would be different: call it
  $\bar{S}$. With that one, the forward would be $F_{t,T} = \bar{S}_t \exp \{(r_{t,T})(T-t)\}$. 
  Given our original index price $S_T$, this motivates us to associate a random
  total-return index $\bar{S}_T \equiv S_T \, \e^{\delta_{t,T}(T-t)}$ which incorporates 
  hypothetically reinvested dividends. The associated log total-return is
  $\bar{X}_T = \log (\bar{S}_T/\bar{S}_t) = X_T + \delta_{t,T}(T-t)$, and the associated risk-neutral density is
  found by a simple translation:   
  
  \be  \mbox{\underline{log total-returns}}: \quad q_{\bar{X}_T}(x) = q_{X_T} \left( x - \delta_{t,T} \, (T-t) \right). 
        \label{eq:divadj)} \eb

  \pbold{Transforming to the real-world probability density.}
    Finally, we move from risk-neutral ($\QBB$-measure) densities to real-world 
    ($\PBB$-measure) densities using the exponential tilt
  associated to a risk-averse representative agent model. 
  The agent, with time-$t$ wealth $W_t$, chooses 
  investments with payoffs at $T > t$ by maximizing her expected utility
  $\bar{U}_{t,T}(W_t) = c^{T-t} \, \Et{(W_T/W_t)^{(1-\kappa)}}$. Here $c \le 1$ is
  an impatience parameter that need not be estimated, and $\kappa \ge 0$ is the agent's
  \emph{Coefficient of Relative Risk Aversion} (CRRA). 
  To clear the market, all the agent's wealth must be optimally invested in the
  aggregate securities market; thus $W_T = \bar{S}_T$, the total return equity index from above. While this utility model is well-known in spirit, there are many variations: our version is developed in Appendix A. Using total return pdf's, it leads to the $\PBB \Leftrightarrow \QBB$ duality:
  
  \be q_{\bar{X}_T}(x) = \frac{\e^{-\kappa x} \, p_{\bar{X}_T}(x)}{\int \e^{-\kappa x}  p_{\bar{X}_T}(x) \, dx} \quad \Leftrightarrow \quad p_{\bar{X}_T}(x) = \frac{\e^{\kappa x} \, q_{\bar{X}_T}(x)}{\int \e^{\kappa x} q_{\bar{X}_T}(x)  \, dx}.           \label{eq:pqduality}  \eb

  \subsection{Calculating the ERP} Equation (\ref{eq:ERPdef}) uses RWD's; specifically, the $\PBB$-total-return densities. With    (\ref{eq:pqduality}), we have 
  
  \be   \mbox{ERP}_{t,T}  = \frac{\int \e^{(1+ \kappa) x} \, q_{\bar{X}_T}(x) \, dx}
         {\int \e^{\kappa x}  \, q_{\bar{X}_T}(x)   \, dx} - (1+R^f_{t,T}). \label{eq:ERP2}  \eb    
 Use (\ref{eq:GMM}),(\ref{eq:divadj)}) and some routine calculations to evaluate (\ref{eq:ERP2}). 
 The result for the option-based ERP is relatively simple -- on an annualized percent basis:
   
  \begin{empheq}[box=\fbox]{align}
    \mbox{ERP}_{t,T}^{(ann\%)}(\kappa) &= \frac{100}{T-t} \times
      \left\{  \left( \e^{\delta_{t,T} \tau} \sum_{i=1}^N \tilde{w}_i \, \e^{\alpha_i + (\kappa + \frac{1}{2}) v_i} \right) - 
             \e^{r_{t,T} \tau} \right\}, 
          \label{eq:ERPfinal}  \\
    & \mbox{using} \,\, \tau = T-t, \quad \alpha_i = \mu_i \tau, \,\, v_i = \sigma_i^2 \tau,  \nonumber \\      
     & \quad \,\, \gamma_i = \kappa \, \alpha_i  + \Smallfrac{1}{2} \kappa^2 v_i, \,\, \mbox{and}
          \,\, \tilde{w}_i = w_i \e^{\gamma_i}/\sum_{i=1}^N w_i \e^{\gamma_i}. \nonumber
   \end{empheq}  
 Here the $\{w_i,\mu_i,\sigma_i\}$ are the GMM fits to option quotes at $t$ for expiration $T$.
 
 \Pbreak 
 This concludes our summary.        
  
  \newpage

  \section{Estimation of the CRRA parameter $\kappa$} \label{sec:kappaestimates}
  
  There is a large literature on this topic. A classic study is \cite{friend:1975} who concluded (adapting to our notation):
  
  \begin{quote}
  	The empirical results ... indicate that the assumption of constant proportional risk aversion for households is
  	a fairly accurate description of the marketplace ... The implication is that $\kappa$ for the typical household
  	is in excess of 1.0 -- contrary to the proportion of the \emph{log} utility function. Since the market price of
  	risk is probably around 2.0 or more, $\kappa$ is more likely to be in excess of two.   
  \end{quote}
 Unfortunately, the subsequent literature muddied up this relatively clean picture.\footnote{\label{ft:bliss} For example,
 	\cite{bliss:2004} present (their Table VII) literature estimates from eleven studies with $\kappa$ estimates
 	ranging from 0-55.} Our approach is straightforward and agreeable with Friend \& Blume: we estimate
    $\hat{\kappa}=3 \pm 0.5$.  At the
    end of this section, we compare with the study mentioned in footnote \ref{ft:bliss}. 
    
    We begin with the martingale relations.
   
   \pbold{Martingale relations.} In a risk-neutral world, the ERP is zero and (\ref{eq:ERPdef}) reads
   
   \be    \Eqt{R^e_{t,T} - R^f_{t,T}} = 0 \label{eq:mart1}  \eb
   for all $(t,T)$ pairs. Recall $1 + R^e_{t,T} = \bar{S}_T/\bar{S}_t$, where $\bar{S}_t$ is
   a total-return index (i.e., including reinvested dividends). Thus, (\ref{eq:mart1}) has the well-known equivalent:   
   \be   \bar{S}_t = \frac{1}{1 + R^f_{t,T}}  \Eqt{\bar{S}_T}.   \label{eq:mart2}  \eb 
   Recall $1/(1 + R^f_{t,T}) = D_{t,T}$ in our previous notation. Thus, (\ref{eq:mart2}) 
  says the discounted (total-return) index behaves like a martingale under the risk-neutral
  measure -- the starting point for many dynamical stochastic process models. This justifies calling either  (\ref{eq:mart1}) or  (\ref{eq:mart2}) `martingale relations'.
  
  \pbold{The estimator.}
  To estimate $\kappa$, consider $R^e_{t,T}$ where $T-t$ is, for example, one-month. Then employ (\ref{eq:mart1}) 
  on an unconditional basis using a long series of historical one-month equity total returns.  
  That is, first write (\ref{eq:mart1}) as 
  
  \be  \int e^{-\kappa x_t} (R^e_t - R^f_t) \, p(x_t) \, d x_t = 0.       \label{eq:mart1alt} \eb
  Here  $R^e_t = \e^{x_t} -1$ and   $p(x_t)$ is the unconditional density for $x_t$, the month-$t$ 
   log-total-return of the equity aggregate. 
  The $R^f_t$ are then the monthly Tbill returns. Next, estimate $p(x_t)$ from the empirical density using 
   $\{x_i = x_{t(i)}: i=1, \cdots, M\}$ a list of $M$ monthly historical returns from (let's say) 1926 to date. 
   (We consider other periods also).  The  empirical density is $p(x) = \frac{1}{M} \sum_{i=1}^M \delta(x - x_i)$, using the Dirac delta.
   Upon substitution in (\ref{eq:mart1alt}): $\kappa$ is estimated by
   $\hat{\kappa}$, the solution to
   
   \begin{empheq}[box=\fbox]{align}   
   f(\hat{\kappa}) = 0,\quad \mbox{where} \quad f(\kappa) = \sum_{i=1}^M \e^{-\kappa x_i}  (\e^{x_i} - 1 - R^f_i).
    \label{eq:kappaestimator} 
    \end{empheq}
   In other words, $\hat{\kappa}$ is the exponential tilt parameter that turns the historical equity 
   excess return sequence into a zero-mean sequence. This is `dynamics-free':  no
   specific stochastic process is imposed upon the sequence. Of course, our approach relies upon the assumption that
   this exponential change of measure is indeed used by the market (or the representative agent if you like). 
   Given the need to either impose dynamics or impose a preference model, the latter choice is the 
   minimal (``model-light") one.

  \subsection{Data sources for historical equity returns}
  
  I use two equity indices as proxies for the aggregate U.S. equity market:
  
  \begin{itemize}
  \item[(i)] The S\&P 500/Composite Index (SBBI);
  \item[(ii)] All US exchange-listed stocks (Fama and French online data from CRSP)
  \end{itemize}
Some additional detail:
  
  \pbold{S\&P 500 Index.} The S\&P 500 Index has modern
  form (close to 500 stocks) starting March 4, 1957. However,
  this series is commonly joined with the earlier ``S\&P Composite" (90 large U.S. stocks),
  extending back to 1926 and still called the S\&P 500.  Our main source for the joined series
   was the ``SBBI Yearbook: Stocks, Bonds, Bills, and Inflation" \cite{SBBI:2016}.  I used
  	SBBI data for monthly total returns through calendar year 2017 and updated the results
  	myself through June 2019. For the update (capital appreciation), I used Mathematica's built-in curated data ({\MM{FinancialData["SP500"]}}).
  	To update the dividend income returns, I used the ``S\&P 500 Dividend Points Index (SPXDIV)",
  	available online at {\url{https://us.spindices.com/indices/equity/sp-500-dividend-points-index-quarterly}}.
  	There was excellent agreement \newline among these various sources where they overlapped.

  \pbold{Fama and French data.} While SBBI data is monthly, Kenneth French provides (and updates online) a daily series used
  by him and Eugene Fama in their research. (FF data for short). Specifically, they provide $R^e_t - R^f_t$ a capitalization
  weighted, daily excess return on the `market', represented by ``all CRSP firms
  incorporated in the US and listed on the NYSE, AMEX, or NASDAQ that have
  a CRSP share code of 10 or 11". CRSP is the Center for Research in Security
  	Prices, part of the Booth School of Business at the University of Chicago. The Ken French data library and further details
  	may be found at \url{https://mba.tuck.dartmouth.edu/pages/faculty/ken.french/data_library.html}. 
  The risk-free rate $R^f_t$ is described as the ``Treasury bill rate (from Ibbotson Associates)", likely the SBBI series prorated to daily returns.
  
  \subsection{S\&P500 results}
  Table \ref{tab:kappaSBBI} shows the $\hat{\kappa}$ estimates from the monthly SBBI series using
  various start dates through June 2019.  Also shown are various moments of the true series and the inferred risk-neutral distribution (from tilting at the estimated $\hat{\kappa}$).
  Recall the `excess returns' are the (annualized) $R^e_t - R^f_t$ series; those  have zero $Q$-means by construction.  Higher moments shown are computed from log-returns: $\log(1 + R^e_t)$. 
  
  Why those starting dates? Jan 1926 starts the entire series. Jan 1950 roughly
  starts the post-WWII period, often considered a structural break to a world of reduced
  average volatility and less frequent recessions. Recall Apr 1957 marks the start of the `true' S\&P 500 Index. Finally, Jan 1987 and Jan 1988 are convenient starts to show both pre- and post- Oct 19, 1987 data, as the Black Monday market crash is often considered an outlier.
  
  As one sees from Table \ref{tab:kappaSBBI}, $\hat{\kappa}$ estimates all lie within a range of $2-4$. Post-WWII $\kappa$ estimates tend to be higher than the entire series. Those are 
  the results from that table important for the ERP.\footnote{  
  Some incidental observations. The risk-neutral standard deviations are always larger than the true values --  by amounts ranging from roughly 1 to 1.5 vol points (annualized percentage points). Risk-neutral skewness's are uniformly more negative than real-world. Kurtoses are smaller post-WWII and have similar risk-neutral and real-world values for every start date.}

  \begin{table}[t] 
  	\caption{{\bf{S\&P 500: CRRA $\mathbf{\hat{\kappa}}$ estimates and associated statistics.}}\newline 
  		Various start dates through June 2019. Monthly observations. Means and
  		Std Dev's are annualized and in percent. For example, `Mean excess returns' are
  		$12 \, \times 100 \, \times$ (avg monthly total returns less monthly Tbill returns). 
  		Standard deviations, skewness, and kurtosis are
  		based upon continuously compounded total returns (log-returns). `True' =
  		actual: realized. `RN'= risk-neutral: the exponentially-tilted empirical $Q$-distribution
  		with tilt parameter $\hat{\kappa}$.}
  	\begin{center}
  		\begin{tabular}{cccrlrlrlrlrl} 			
  			\toprule 
  			      &  & & \multicolumn{2}{c}{Mean}  &   \\
  			Start &  & &\multicolumn{2}{c}{Excess Return} &\multicolumn{2}{c}{Std Dev} 
  			&\multicolumn{2}{c}{Skewness} &\multicolumn{2}{c}{Kurtosis} \\	                
  			month & $N_{obs}$ & $\hat{\kappa}$ & True & RN & True   &  RN    & True & RN  & True & RN \\		
  			\cmidrule(r){1-1}   \cmidrule(r){2-2} \cmidrule(r){3-3} \cmidrule(r){4-5}  \cmidrule(r){6-7}   \cmidrule(r){8-9}  \cmidrule(r){10-11} 
  			Jan 1926  & 1122 & 2.28  & 8.14 & 0 & 18.6 & 19.7 & -0.50 & -1.29 & 11.0 & 10.4  \\
  			Jan 1950  & 834  & 3.58  & 7.73 & 0 & 14.3 & 15.2 & -0.67 & -0.95 & 5.5 & 6.3 \\
  			Apr 1957  & 747  & 2.96  & 6.52 & 0 & 14.5 & 15.3 & -0.69 & -0.93 & 5.6 & 6.2 \\
  			Jan 1987  & 390  & 3.33 & 8.04 & 0 & 15.0 & 16.5 & -1.09 & -1.36 & 6.7 & 7.4    \\
  			Jan 1988 & 378  & 3.83 & 8.15 & 0 & 14.2 & 15.2 & -0.78 & -0.90 & 4.7 & 4.8 \\
  			\bottomrule 
  		\end{tabular}
  	\end{center}
  	\label{tab:kappaSBBI}
  \end{table}

  \subsection{Temporal aggregation of the S\&P500 data}
  Strictly speaking, the stock market never offers a stationary `return generating process', in the sense of
  a well-specified casino game. Nevertheless, if the risk-aversion model is
  not too far off the mark, given long `pseudo-stationary' samples, one
  would expect similar $\hat{\kappa}$ estimates regardless of the return observation
  frequency: daily, monthly, quarterly, and so on. Given our monthly S\&P 500 total returns,
  we aggregate the data into longer periods and repeat the estimating procedure.
  
  \pbold{Results.}  
  In Table 	\ref{tab:kappaSBBIaggregation} we show the effect of this temporal aggregation, using 
  the longest SBBI data period: Jan 1926 -- June 2019. When aggregating, there are
  two choices: overlapping or non-overlapping periods. We show both choices -- with the
  exception of yearly. That's because, except for years, the non-overlapping
  data still end exactly on June 30, 2019. Note: we also call the data frequency length 
  the `horizon length'.
  
  As one sees, the
  $\hat{\kappa}$ estimates are not very sensitive to aggregation: that's evidence in favor of the risk-adjustment model. However, there is a small tendency for $\hat{\kappa}$ to decrease with the horizon length.\footnote{More incidental observations on Table 	\ref{tab:kappaSBBIaggregation}. 
  True standard deviations are seen fairly constant under aggregation, but risk-neutral
  standard deviations are growing with the aggregation horizon. This is consistent with a 
  \emph{term structure} to volatility risk premiums, which also tends to increase with horizon
  length. For example, there is an unconditional term structure to VIX's which has a similar increase with horizon lengths.  
  Risk-neutral skewness seem fairly stable with the horizon length. Both the true
  and risk-neutral kurtosis are seen decreasing with the horizon length.}

  \begin{table}[t] 
  	\caption{{\bf{S\&P 500: CRRA $\mathbf{\hat{\kappa}}$ estimates Jan 1926 -- June 2019.}}\newline 
  		Various observation frequencies with means and std devs annualized.}
  	\begin{center}
  		\begin{tabular}{ccccrlrlrlrlrl} 			
  			\toprule 
  		   	     &       &  &  & \multicolumn{2}{c}{Mean}  &   \\
  			Data & Over- &  & &\multicolumn{2}{c}{Excess Returns} &\multicolumn{2}{c}{Std Dev} 
  			&\multicolumn{2}{c}{Skewness} &\multicolumn{2}{c}{Kurtosis} \\	                
  			frequency & lap?& $N_{obs}$ & $\hat{\kappa}$ & True & RN & True   &  RN    & True & RN  & True & RN \\		
  			\cmidrule(r){1-1}   \cmidrule(r){2-2} \cmidrule(r){3-3} \cmidrule(r){4-4}
  			\cmidrule(r){5-6}  \cmidrule(r){7-8}   \cmidrule(r){9-10}  \cmidrule(r){11-12} 
  			monthly & no & 1122 & 2.28  & 8.14 & 0 & 18.6 & 19.7 & -0.50 & -1.29 & 11.0 & 10.4  \\
  			quarterly&no & 374  &  1.97 & 8.66 & 0 & 20.6 & 21.9 &  0.06 & -1.05 & 10.9 & 7.5 \\
  			quarterly&yes& 1120 & 2.10  & 8.39 & 0 & 19.4 & 21.5 & -0.35 & -1.43 & 11.0 & 8.7\\
  			6-months&no  & 187  & 1.95  & 8.55 & 0 & 19.8 & 23.5 & -0.91 & -1.33 & 6.8 & 5.9 \\
  			6-months&yes & 1117 & 2.05  & 8.47 & 0 & 19.2 & 22.9 & -0.92 & -1.33 & 6.7 & 6.0 \\ 
  			yearly  &yes & 1111 & 1.82  & 8.89 & 0 & 20.1 & 26.7 & -1.08 & -1.49 & 6.9 & 6.2 \\ 
  			\bottomrule 
  		\end{tabular}
  	\end{center}
  	\label{tab:kappaSBBIaggregation}
  \end{table}

  \begin{table}[th] 
  	\caption{{\bf{FF data: CRRA $\mathbf{\hat{\kappa}}$ estimates and associated statistics.}}\newline 
  		Data: all US-incorporated stocks listed on the NYSE, AMEX, and NASDAQ exchanges. Various start dates 
  		through Apr 30, 2019. Means and Std Dev's are annualized and in percent. Daily mean excess returns are
  		$252 \, \times 100 \, \times$ (avg daily total returns less Tbill returns). 
  		Standard deviations, skewness, and kurtosis are
  		based upon continuously compounded total returns (log-returns). `RN'= Risk-neutral:
  		the exponentially-tilted empirical $Q$-distribution.}
  	\begin{center}
  		\begin{tabular}{lccrlrlrlrlrl} 
  					\toprule 
  			 I. Daily &  & & \multicolumn{2}{c}{Mean}  &   \\
  			 &  & &\multicolumn{2}{c}{Excess Returns} &\multicolumn{2}{c}{Std Dev} 
  			&\multicolumn{2}{c}{Skewness} &\multicolumn{2}{c}{Kurtosis} \\	                
  			Start & $N_{obs}$ & $\hat{\kappa}$ & True & RN & True   &  RN    & True & RN  & True & RN \\		
  			\cmidrule(r){1-1}   \cmidrule(r){2-2} \cmidrule(r){3-3} \cmidrule(r){4-5}  \cmidrule(r){6-7}   \cmidrule(r){8-9}  \cmidrule(r){10-11} 
  			Jul 1, 1926  & 24473 & 2.58  & 7.40 & 0 & 16.9 & 17.0 & -0.43 & -0.92 & 20.5 & 22.5  \\
  			Jan 1, 1950  & 17530 & 3.48  & 7.68 & 0 & 14.8 & 15.1 & -0.84 & -1.52 & 22.5 & 30.3 \\
  			Apr 1, 1957  & 15627 & 2.85  & 6.69 & 0 & 15.3 & 15.5 & -0.80 & -1.34 & 21.8 & 27.8 \\
  			Jan 1, 1987  & 8147  & 2.67  & 8.49 & 0 & 17.7 & 18.1 & -0.95 & -1.52 & 21.6 & 26.8    \\
  			Jan 1, 1988  & 7894  & 2.93  & 8.75 & 0 & 17.2 & 17.4 & -0.35 & -0.61 & 11.4 & 11.5 \\
  			\bottomrule 
  			\\
  			\\	
  			 			\toprule 
  			 II. Monthly &  & & \multicolumn{2}{c}{Mean}  &   \\
  			 &  & &\multicolumn{2}{c}{Excess Returns} &\multicolumn{2}{c}{Std Dev} 
  			&\multicolumn{2}{c}{Skewness} &\multicolumn{2}{c}{Kurtosis} \\	                
  			Start & $N_{obs}$ & $\hat{\kappa}$ & True & RN & True   &  RN    & True & RN  & True & RN \\		
  			\cmidrule(r){1-1}   \cmidrule(r){2-2} \cmidrule(r){3-3} \cmidrule(r){4-5}  \cmidrule(r){6-7}   \cmidrule(r){8-9}  \cmidrule(r){10-11} 
  			Jul 1, 1926  & 1114 & 2.27   & 7.96 & 0 & 18.4 & 19.5 & -0.56 & -1.23 & 9.9 & 9.6  \\
  			Jan 1, 1950  & 832  & 3.39   & 7.73 & 0 & 14.7 & 15.7 & -0.77 & -1.06 & 5.8 & 6.6 \\
  			Apr 1, 1957  & 745  & 2.83   & 6.68 & 0 & 15.0 & 15.9 & -0.78 & -1.01 & 5.7 & 6.4 \\
  			Jan 1, 1987  & 388  & 3.19   & 8.11 & 0 & 15.4 & 17.0 & -1.21 & -1.48 & 7.2 & 7.9    \\
  			Jan 1, 1988  & 376  & 3.72   & 8.34 & 0 & 14.6 & 15.7 & -0.87 & -0.97 & 4.8 & 4.9 \\
  			\bottomrule 			
  		\end{tabular}
  	\end{center}
  	\label{tab:kappaFF}
  \end{table}

   \subsection{Fama and French data results}
  
  To enable  a direct comparison with SBBI monthlies, we also show FF monthlies.  All FF series begin on July 1, 1926 and, as of this writing, have been updated through April 30, 2019.
  We repeat the analysis from Table \ref{tab:kappaSBBI} with results now found in Table \ref{tab:kappaFF}.

  \pbold{Results.} For Panel I(FF Daily), the $\hat{\kappa}$ estimates are broadly consistent with Table \ref{tab:kappaSBBI}; however, they now have a narrower range: 2.5 - 3.5. 
  For Panel II (FF Monthly), the $\hat{\kappa}$ estimates are quite close to
  Table \ref{tab:kappaSBBI} (SBBI/S\&P 500) with a range 2.3 - 3.7.\footnote{
  Incidental observations on the FF results. Re Panel I: the RN standard deviations are
  now only slightly larger than the true ones. Given now daily observations, this is
  additional evidence of a monotonically increasing term structure for the (annualized) unconditional risk-neutral  volatility $\sigma_Q(\tau)$, associated to returns with horizon length $\tau$. Recall the discussion  for Table \ref{tab:kappaSBBIaggregation}.
   The daily return RN skewness are found in the range -0.6 to -1.5.
  Daily kurtoses are, unsurprisingly, now much larger than seen in Table \ref{tab:kappaSBBIaggregation}
  for periods containing the Oct 1987 crash. RN kurtoses tend to be larger than the real-world. All kurtoses fall significantly post-crash.
  
  Re Panel II: As with the SBBI monthlies, the RN standard deviations are
  now about 1-1.5 vol points (1-1.5\%) larger than the true ones. Again, this is a term structure effect.   The monthly return RN skewness are found in the range -1 to -1.5.
  The kurtosis pattern is similar to Table \ref{tab:kappaSBBI}.}

  \newpage
  
  \subsection{Summary and contrast with related literature} 
  In summary, based upon both the SBBI and FF results, we adopt the central estimate $\hat{\kappa}=3$, with a confidence range of $2.5 - 3.5$ for US equities. 
  
  \Pbreak
  An interesting and related study is \cite{bliss:2004}. Among other things, the authors use a similar
  power utility model to transform an estimated risk-neutral density from S\&P 500 (futures) options
  to a real-world density. With different data and methods than ours, they argue for a horizon-dependent
  coefficient of relative risk aversion: $\kappa(\tau)$ (in our notation) where $\tau= T-t$. They estimate  $\hat{\kappa}(\tau)$ for various horizons by maximizing the forecast ability of the
  corresponding inferred real-world densities over 1983-2001. Their estimates show a significant horizon effect: declining from $\hat{\kappa} \sim8$ at a one-week horizon to $\hat{\kappa} \sim2$ at a 6-week horizon, 
  their upper limit (Table V in their paper). 
  
  \Pbreak  
  In contrast, our evidence supports a reasonably constant $\kappa$ with a smaller horizon dependence. We investigated the effect of temporal aggregation (the horizon length) on our estimates. And recall we indeed found a tendency for $\hat{\kappa}$ to decline with the horizon; however, the horizon effects we saw were much smaller and sometimes in the opposite
  direction: recall Table \ref{tab:kappaSBBIaggregation} (S\&P 500)  and Table \ref{tab:kappaFF} (CRSP all US equity). In general, the horizon differences we saw should be considered subsumed by our overall estimated uncertainty: a `true' $\kappa$ lying somewhere in 2.5-3.5.

 \newpage
  
  \section{Option data and handling}
  I acquired end-of-day SPX option data for all Wednesdays from Jan 2018 through June 2019 from the CBOE’s LiveVol service: ``End-of-Day Option Quotes with Calcs". These Wednesdays are the `trade dates'. Actually `end-of-day' is
  a slight misnomer: the files record option quotes and CBOE-calculated option implied volatilities (IV’s) at 15:45 New York time. This time is 15 minutes prior to the regular stock and option market session close in NYC and Chicago. According to the CBOE:
  \begin{quote}
  ``Implied volatility and Greeks are calculated off the 1545 time-stamp, considered a more accurate snapshot of market liquidity than the end of day market".
  \end{quote}
  I selected three trade dates for analysis, hopefully reported here in enough detail to encourage
  replication studies:
  
  \begin{itemize}
  	\item[(i)] Feb 7, 2018: a relatively high volatility environment, two days after the `Volpocalyspse'.\footnote{The
  		Feb 5, 2018 volatility event, more amusingly the `Volpocalypse', is quite interesting in its own right.
  	    On that day the Dow Jones Industrial Average lost 1175 points, its worst point decline ever.
        However, the percentage loss was only 4.6\%, certainly not a crash, and SPX only lost 4.1\%. Given only those facts, one might guess that VIX would rise about 20\% -- again nothing to write home about. Instead, VIX rose almost
        100\%, triggering a `termination event' in a popular exchange-traded fund (ticker XIV). This (short) volatility product,
        by design, maintained a short position in VIX futures: about \$1.5 billion worth going into the session (NAV/share 
        $\approx$ \$100). Of course, if you
    are short \$X in a collateralized future, and the future rises 100\%, you will lose your \$X. Indeed, XIV ended the day with
    an NAV around \$60 million (NAV/share = \$4.22), losing about 95\%. As provided by its prospectus, under any daily loss exceeding 80\%, the sponsor (Credit Suisse) could, if it so chose, close the product: the fund was indeed closed. 
    
       While the fund worked `correctly' (i.e., as described in its prospectus), investors were correct to be surprised
       by VIX doubling under a 4\% one-day loss in the SPX. I am reminded of a famous dictum from physicist Murray Gell-Mann,
       recently deceased: ``everything that is not forbidden is mandatory".}   
  	\item[(ii)] Aug 8, 2018, the date of a local low in VIX. 
  	\item[(iii)] June 26, 2019, last date of my data set, with VIX in the mid-teens.
  \end{itemize} 
   The motivation was to compare ERP term structures in relatively high, low, and middling volatility environments as measured by
   VIX: see Fig. \ref{fig:VIXplot}.

  \subsection{Preprocessing} The raw CBOE files come one per trade date. For a given trade date, they first needed to be sorted into separate files for each root symbol and expiration. Also, cost-of-carry
  parameters for each expiration need to be identified. These two tasks form the preprocessing step.
  
  \Pbreak
  S\&P 500 index options are cash-settled, European-style,
  options of two types:
  
  \begin{itemize}
  	\item ``AM" options with root SPX, and
  	\item ``PM" options (Weeklys) with root SPXW. 
  \end{itemize}
 SPX (am) options were the first to be introduced and expire on traditional third Fridays of each month. They cash settle based upon a special SPX quotation computed at the opening of the expiration trading day. SPXW (pm) options expire on a variety of weekdays (including those third Fridays); they cash settle based upon the end-of-day closing SPX index value (4:00pm New York time). This is the current root symbology and applies to the data used in this article. (Prior to May 2017, the S\&P 500 root symbology was somewhat different). 
  
  \Pbreak  
  As it turned out, this first sorting resulted in 42 separate expirations for each of the three trade dates analyzed.
  For example, see the first two columns of Tables \ref{tab:ERP020718} and \ref{tab:ERP080818} for the roots and expirations
  for the Feb 7, 2018 and Aug 8, 2018 trade dates.

  \subsubsection{Cost of carry: the VIX white paper method} \label{sec:VIXwp} 
  There are various ways to estimate  $(r,\delta)$. For the data shown here, we used the ``VIX white
  paper method".\footnote{\url{https://www.cboe.com/micro/vix/vixwhite.pdf}} In this method, the riskless
  rate $r_{t,T}$ is taken to be a US Treasury yield for the same maturity $T$. More specifically, starting from
  the Daily Treasury Yield Curve rates available for each trade date at the US Treasury's web site, one can
  interpolate a value for $r_{t,T}$. Next, one determine the forward SPX level, $F$, by identifying the strike price
  $K_{*}$ at which the absolute difference between the call and put prices, $(C,P)$, (using the bid-ask quote average) is smallest. From those, an option-implied forward price, $F = K_{*} + \e^{r_{t,T}(T-t)} (C-P)$, is calculated. Finally, writing 
  $F  = F_{t,T} = S_t \, \e^{(r_{t,T}-\delta_{t,T})(T-t)}$, where $S_t$ is the 15:45 trade date index value, one infers a value for the dividend yield $\delta_{t,T}$. 
  
 \Pbreak
  While the VIX white paper method is attractive, it may not be the best estimator of what option market makers
  are actually experiencing for rates and yields. For example, for option  maturities greater than 3 months,  
  it tends to produce dividend yields that are
  low relative to projections based upon historical dividends. The reason for that may be that the
  US Treasury rate (even though a `term' rate) is low relative to the typical funding/investing rates 
  that are paid/received by market makers and other professional traders. 
  To clarify the issue, a second cost-of-carry method was considered, using a regression based upon
  put-call parity. While that alternative method indeed resulted in higher $(r,\delta)$'s, the option-implied
  forward prices were quite close. As a result, the ERP estimates were also quite close under the two methods. 
  Details are found in Sec. \ref{sec:altcoc}.
  
  \newpage
  
  \section{The GMM: option values and fitting} \label{sec:OIERP}  
  
  \subsection{Option values} \label{sec:Optvals}
  At each trade date $t$, there are typically 50-300 options expiring at each expiration $T$. Each of these options has a  trade time 15:45 bid (which may be zero) and an ask. The mid-quote is the average of
  the bid and ask; we fit mid-quotes for out-of-the-money options: puts for $K < S_t$ and calls for
  $K \ge S_t$, where $K$ is the option strike price. After some filtering to remove zero-bid quotes (and sometimes larger bid quotes), the number of options actually fitted, $N_{opts}$, is shown by the corresponding column in Tables \ref{tab:ERP020718}-\ref{tab:ERP080818}.
  
  An $N$-component Gaussian mixture model (GMM) is a weighted sum for the $\QBB$-measure (risk-neutral) density
  for $X_T = \log S_T/S_t$, given at (\ref{eq:GMM}). Let's write  (\ref{eq:GMM}) as \newline
  $q_{X_T}(x) = \sum_{i=1}^N w_i \, \phi(x;\mu_i (T-t),\sigma_i^2 (T-t))$, where $\phi(x;\mu,v)=
  \exp(-(x-\mu)^2/2 v)/\sqrt(2 \pi v)$. With discount factor $D_{t,T} = \exp(-r_{t,T} \tau)$, 
  using $\tau \equiv T-t$, call values $C_{t,T}$ are given by   
  \be C_{t,T}(S_t,K) = D_{t,T} \, \Eqt{(S_T-K)^+} = D_{t,T} \sum_{i=1}^N w_i \, C_i. \label{eq:Cval} \eb
  In (\ref{eq:Cval}), \emph{undiscounted} component call values $C_i$ are found from routine calculations:  
  \be C_i = S_t \, \e^{(\mu_i + \frac{1}{2} \sigma_i^2) \tau} \Phi(d_{i,1}) - K \Phi(d_{i,2}), \label{eq:callcomponent} \eb
  \[ \mbox{using} \quad d_{i,1} = \frac{ \log \frac{S_t}{K} + \mu_i \tau}{\sigma_i \sqrt{\tau}} + \sigma_i \sqrt{\tau},
   \quad \mbox{and} \quad d_{i,2} = d_{i,1} - \sigma_i \sqrt{\tau}. \]   
   Similarly, for put values: $P_{t,T}(S_t,K) = D_{t,T} \Eqt{(K-S_T)^+} = D_{t,T} \sum_{i=1}^N w_i \, P_i$, where
   \be P_i = K \Phi(-d_{i,2}) - S_t \, \e^{(\mu_i + \frac{1}{2} \sigma_i^2) \tau} \Phi(-d_{i,1}). \label{eq:putcomponent} \eb
  Free parameters $\{w_i,\mu_i,\sigma_i\}$ are chosen to minimize an objective function subject to two constraints:
   (i) the norm condition: $\sum_{i=1}^N w_i = 1$, and  
  \be  \mbox{(ii) the martingale condition}: \quad \sum_{i=1}^N w_i \, \e^{(\mu_i + \frac{1}{2} \sigma_i^2) \tau} = \e^{(r_{t,T}-\delta_{t,T}) \tau} \label{eq:martcond} \eb 
 With the constraints, 
  the GMM  satisfies put-call parity, the model-independent relation:
  \be C_{t,T} - P_{t,T} = S_t \, \e^{-\delta_{t,T} \tau} - K \, \e^{-r_{t,T} \tau}
     = \e^{-r_{t,T} \tau} (F_{t,T} - K).  \label{eq:putcallparity} \eb
  
  \pbold{Discussion.} Three comments:
  \begin{itemize}
  	\item[(i)] Consider the model-independent forward price $F_{t,T} =  \Eqt{S_T} = S_t \, \e^{(r_{t,T}-\delta_{t,T}) \tau}$,
   which can be interpreted as the value of the undiscounted, zero-strike call option. 
   With the GMM (\ref{eq:Cval}), define \emph{component} forward prices $F^i_{t,T}$
   as the undiscounted, zero-strike, component call values. From (\ref{eq:callcomponent}),
    $F^i_{t,T} = S_t \, \e^{(\mu_i + \frac{1}{2} \sigma_i^2) \tau} $. Thus, an equivalent version
    of the martingale condition (\ref{eq:martcond}) is $\sum_{i=1}^N w_i F^i_{t,T} = F_{t,T}$.
  \item[(ii)]  As one check, with $N=1$, $w_1=1$ and the martingale condition is $\mu_1+ \frac{1}{2} \sigma_1^2 = r_{t,T}-\delta_{t,T}$. With that, as expected, (\ref{eq:Cval}) reduces to the Black-Scholes call option formula. 
  \item[(iii)] As another check, consider the ERP at (\ref{eq:ERPfinal}) for a risk-neutral agent: $\kappa=0$.
  With the martingale condition: $\mbox{ERP}_{t,T}^{(ann\%)}(\kappa=0)= 0$, as expected. 
  \end{itemize}

   \pbold{Time measurements.} Notice that column 2 in Tables \ref{tab:ERP020718}-\ref{tab:ERP080818} shows the integer number of calendar days from the trade date to the expiration
  date of the options. When times to expirations, $T-t$, are needed in option formulas such as (\ref{eq:callcomponent}), 
   more precise time measurements were used.
  
  For example, the first entry in Table \ref{tab:ERP020718} shows a
  PM (SPXW) option with 2 days to expiration. Then, $T-t$, which is measured in years, was taken to be
  $(2 + 0.25/24)/365$, accounting for option quotes $1/4$-hour before the
  regular session close. Similarly, the 4th entry in the same table shows an AM (SPX) option with 9 days to
  expiration. For that one $T-t$ was taken to be   $(9 + (0.25 - 6.5)/24)/365$, also accounting for
  the 6.5 hours from the opening to the close of the regular trading session.\footnote{\label{ft:bustime}While our calendar time measurement seems precise, there remains the possibility of a ``business time vs. calendar time" issue. In the tables, one sees the many `dual' expirations with both AM and PM options expiring.
  	The ERP's are reasonably close, given noisy/incomplete data and imperfect GMM fits. But, there might be distortions in the ERPs (remember they are annualized) from the time measurement used here. The most extreme case would be a trade date only one or two days from a dual expiration. While this possibility is not realized in our data, if present, it might prompt some further time adjustment. Complicating any putative business time adjustment would be the need to remove calendar spread arbitrage violations,
  	the topic of Sec. \ref{sec:calarb}.}

  \subsection{Fitting methodology} \label{sec:fitting}
  I employed two goals, seeking to achieve ``market-consistent" fits. First, free parameters were fit by
  minimizing a smooth objective function: (\ref{eq:nominalobj}) below. It uses 
  a particular average deviation of GMM model prices
  from mid-point quotes (using out-of-the-money puts and calls). Call that the ``primary or nominal"
  objective function. Specifically, using $C_i$ to indicate either a put or call, I gave the optimizer the problem:
  
  \be \mbox{Primary objective:} \quad \min_{\{\vec{w},\vec{\mu},\vec{\sigma}\}} \frac{1}{N_{opts}} \, \sum_{i=1}^{N_{opts}} \frac{(C^{mkt}_i - C^{model}_i)^2}{C^{mkt}_i},  \label{eq:nominalobj} \eb 
  subject to the norm and martingale conditions.
  
  Call that objective the `geometric average price error' because the summand represents the geometric average of:
  (i) the square price error $(C^{mkt}_i - C^{model}_i)^2$ and (ii) the relative price error 
   $((C^{mkt}_i - C^{model}_i)/C^{mkt}_i)^2$. To understand this choice, we introduce the
   ``secondary or meta-objective" which was to
   
   \begin{itemize}
   	\item Secondary objective: Find model prices lying within the bid-ask quotes.
   	\end{itemize}
   After some experimentation with various nominal objective functions, I settled on (\ref{eq:nominalobj})
   as a good compromise in light of the secondary objective. For example, if you try to minimize simply the 
   price error, you highly weight close-to-the-money options. This will tend to generate model
   prices for deep out-of-the-money options outside their bid-ask quotes, frustrating the secondary objective.
   Alternatively, minimizing the relative price error over-emphasizes the deep out-of-the-money options at the expense
   of the others. The geometric average is a balanced compromise.    
   
   Ideally, given the nominal objective, one could achieve fits where each model price lay within bid-ask quotes. While this was possible to achieve at many expirations, it proved to be unrealistic to insist on this.
   After all, the data is noisy and the model is imperfect. Instead,
   I adopted the following criteria. First, given a model fit using the primary objective function, I computed
   ``Bid-ask Out statistics" or {\MM{OutStats}} for short. The {\MM{OutStats}} consisted of a two item list: (i) the number of model prices that lay outside the bid-ask quotes, and (ii) the `worst-case' error.
   The worst case error is 0 if all the model prices lay within the bid-ask spread. Otherwise, it
   is the largest absolute price difference between the model and the bid or ask (whichever was closest), and rounded
   to the nearest \$0.01. I also adopted the following qualitative description of the fits:
   
   \newpage
   
   \begin{itemize}
   	\item G (Good): the {\MM{OutStats}} were $(0,0)$ or $(n,\$0.00)$; i.e., no outs, or a worst-case error that 
   	penny rounds to \$0.00;
   	\item A (Acceptable): {\MM{OutStats}} were $(1,err)$ or $(2,err)$; i.e. at most 2 outs with a positive error after
   	penny rounding; 
   	\item W (Weak; likely improvable): all other cases.
   \end{itemize}

   \pbold{Examples.} What do Good and Acceptable fits look like? Figure \ref{fig:BidAskFitPlotG} shows an
   example of a Good fit, with model prices shown as dots within the bid-ask intervals (vertical lines). To better see where
   the model prices are located within the intervals, the inset graph shows expanded detail. In the inset, all 
   prices have been shifted downward by the mid-quote. Thus, if a model price passed exactly through the mid-quote, the
   dot would be at zero in the inset. You can see that all the model prices are close to being centered in
   the quote intervals.
   
   Figure \ref{fig:BidAskFitPlotA} shows an example of an Acceptable fit. Now there is more variation of the
   model prices within the quote interval. Two model prices lie outside the intervals with
   a worst $err=\$0.03$. If you look carefully at the inset (on a monitor), you can see them: one is the 5th strike from the left.
   The other is the 11th strike from the right, \$0.03 below the bid. 
    
  The classification of all of the model fits for Feb 7, 2018 and Aug 8, 2018 is 
  is shown in Tables \ref{tab:ERP020718}-\ref{tab:ERP080818}. For those trade dates, you'll see I  was able to achieve either Good or Acceptable fits for each expiration. The just-discussed
  Figures \ref{fig:BidAskFitPlotG} and \ref{fig:BidAskFitPlotA} are the fit detail for Files 2 and 6, respectively, in Table \ref{tab:ERP020718}. 
  
   For the third trade date examined, 
  there was one expiration with a Weak fit, {\MM{OutStats}}=(3,\$0.01), even after much experimentation. 
  
 \pbold{Experimentation.} We need to briefly discuss the optimizer:
  Mathematica's {\MM{FindMinimum}}. It's a nonlinear local optimizer, which accepts an
  objective function of parameters to fit, subject to their constraints. While somewhat of a black-box, I have found it
  very reliable over many years of use. It implements an Interior Point method and there is much online
  documentation.\footnote{  \url{https://reference.wolfram.com/language/tutorial/ConstrainedOptimizationLocalNumerical.html}}
  Two of the settings are {\MM{PrecisionGoal}} ({\MM{PG}} in tables), and {\MM{MaxIterations}}.
  The optimizer tries to minimize the objective with {\MM{PG}} good digits and within {\MM{MaxIterations}} `steps'. 
  
  Imagine an optimization run using {\MM{PG}}=5, {\MM{MaxIterations}}=250, $N=4$ (Gaussian components),
  and using all out-of-the-money options with non-zero bids. The `Convergence' column in the tables indicates whether or not there was convergence, given {\MM{PG}} and {\MM{MaxIterations}}; if you see `No', then the steps tell you {\MM{MaxIterations}}. 
  In fact, as long as the fit was Good or Acceptable, whether or not  {\MM{FindMinimum}} converged was irrelevant to me. 
  That's because, in addition to having an acceptable fit, the fit error would often be lower \emph{without} convergence.   
  
  But, if the results were a Weak fit, I would experiment with adjustments to the setup. Almost always, this meant  moving to   
   $N=5$  and/or truncating the option set by a few options by boosting {\MM{PutBidMin}}, a
   filter for the minimum put bid allowed.\footnote{On rare occasions, a Weak fit could be improved by
   	moving from $N=5$ to $N=4$. This may seem paradoxical, but remember that the optimizer knows nothing
   	of the Secondary objective.}
    (Including \emph{all} non-zero bids uses {\MM{PutBidMin}}=\$0.05). You can see examples of this last adjustment
   in the tables. Once an adjustment of {\MM{PutBidMin}} had been made, I tried to stick with it
   for other expirations for consistency. (The minimum Call bid was \$0.05 in all optimizations). 
   
   In summary, given a trade date, my method started with a preliminary run through all the expirations. If a fit was Weak, 
   experimentation consisted of tweaking the various setup parameters as explained, until a Good or Acceptable fit had been achieved. This proved to be achievable, with one exception, over $3 \times 42$ expirations. Once these setup parameters were nailed down, a final run was done --- with the final setup and results shown in the tables.

  \newpage

  \section{Results}

  \subsection{Results for densities}
   The fitted risk-neutral densities, $Q(K/S_0)$, real-world densities, $P(K/S0)$, and
   their differences $Q-P$ are plotted in Figs. \ref{fig:pdfsA}-\ref{fig:pdfsC} for
   the Feb 7, 2018 expiration. The vertical axis of each left-most figure has $Q:n$, where
   $n$ is the File number in Table \ref{tab:ERP020718}. This enables you to identify
   the associated option expirations. To save space, I have just plotted the figures for the even file numbers; 
   the odd ones are similar. The real-world densities are estimated using $\kappa=3$.
   
   \pbold{Notable features.} All the densities are smooth and unimodal (single-peaked).\footnote{In
   	principle, the market could face an unusual risk, like a possible asteroid strike or imminent
   	major war, that resulted in a bi-modal density. In practice, single-name equities are the more
   likely place to see bi-modality. There's certainly nothing in general ``no-arbitrage" principles that prevents it.}  
   While the densities are known for $K \in (0,\infty)$, the plots only
   extend from the minimum to maximum strikes used in the fits. The mass coverage under this
   range of strikes is generally very close to one (but slightly less, of course), as suggested by the small values of the densities at their plotted extremes. However, if you look at the furthest expirations, say Q:42, you can
   see the plots do not reach the axis: the mass coverage is relatively smaller with those. 
   
   There is little apparent structure, except that
   one can see some slight ``shoulders" in the furthest expiration Q-densities. It's hard to know if
   those are real features or just fitting artifacts, perhaps connected with the lower mass coverage.
   
   Of course, because of the risk-aversion, the Q-densities place more weight on the downside returns, as shown by the
   difference plots in the third column. 
   
   \subsection{Smile fits and extended smile fits} \label{sec:extsmiles}
   
    The option smile is a plot of the Black-Scholes implied volatility versus the strike price: $IV(K)$.
    There is a market smile and the fitted (GMM) model smile. Fig. \ref{fig:SmilePlots} shows a typical 
    example. The top chart there shows  the market (mid-quote) IV's for the marketable strikes (dots) and $IV(K)$ for the  GMM fit (smooth curve). The bottom chart extends the curve to much smaller and larger strikes -- giving the full picture of the model $IV(K)$. I'll explain the model results and contrast them with related approaches from the literature.
    
    By construction, the GMM RND has Gaussian tails. Gaussian tails imply $IV(K)$ is ultimately `flat' as $K \ra 0$
    or $K \ra \infty$. This will be true under GMM's or spline fits with Gaussian extensions.
    However, approaches may differ in `near' or `far' extension behavior. Far behavior refers to the
    ultimate IV asymptote(s). Near behavior refers to the shape of the extended $IV(K)$ for $K$ close to $K_{min}$ and 
    $K_{max}$ (the smallest and largest marketable strikes).

   The GMM fit yields  $\sigma_{max} \equiv \max_i \{\sigma_i\}$, the largest of the fitted component $\sigma$'s
   from Sec. \ref{sec:Optvals}. Typically, 
   $\sigma_{max} > IV(K_{min})$ (but not too much larger), where $IV(K_{min})$ is the largest IV in the fitted data set.
   It's easy to show that $IV(K)$ ultimately approaches $\sigma_{max}$ for \emph{both} small and large strikes:
   \emph{there is a common flat asymptote}. But, because the SPX smile has such a large skew: $IV(K_{max}) \ll IV(K_{min})$. The  result is that,  near $K_{max}$, the near extension is much more akin to an $IV(K)$ \emph{slope} extension than a flattening: 
   again see Fig. \ref{fig:SmilePlots}. Consequently, while the far extension (the asymptote) is indeed horizontal, the approach is very slow in moneyness terms -- certainly for the large strikes.
   
   Those characteristics differ from the Table \ref{tab:RNDmethods} methods with Gaussian extensions, which have $IV(K)$ near $K_{max}$ close to $IV(K_{max})$.   
   For example, \cite{bliss:2004} and \cite{malz:2014} use flat $IV(K)$ extensions very close
   to $IV(K_{min})$ and $IV(K_{max})$. Thus, these alternative methods yield two \emph{different} asymptotes  and these asymptotes are attained
   in the near region. 
   
   In my opinion, both the $IV(K)$ slope matching the market, and the slow approach of $IV(K)$ to 
   asymptotic flatness are attractive features of the GMM fit. 
   
    \subsection {Results for ERP's} \label{sec:ERPresults}
\begin{figure}[hp] 
  	\caption{{\bf{Equity Risk Premium Term Structure: \newline \hspace*{40pt} 2 days after the ``Volpocalypse" (VIX=27.73)}}.
  	  \newline \hspace*{40pt} (Bands computed with CRRA $\kappa = 3 \pm 0.5$). }
  	\vspace{10pt}
  	\begin{center}
  		\includegraphics[width=0.8\textwidth]{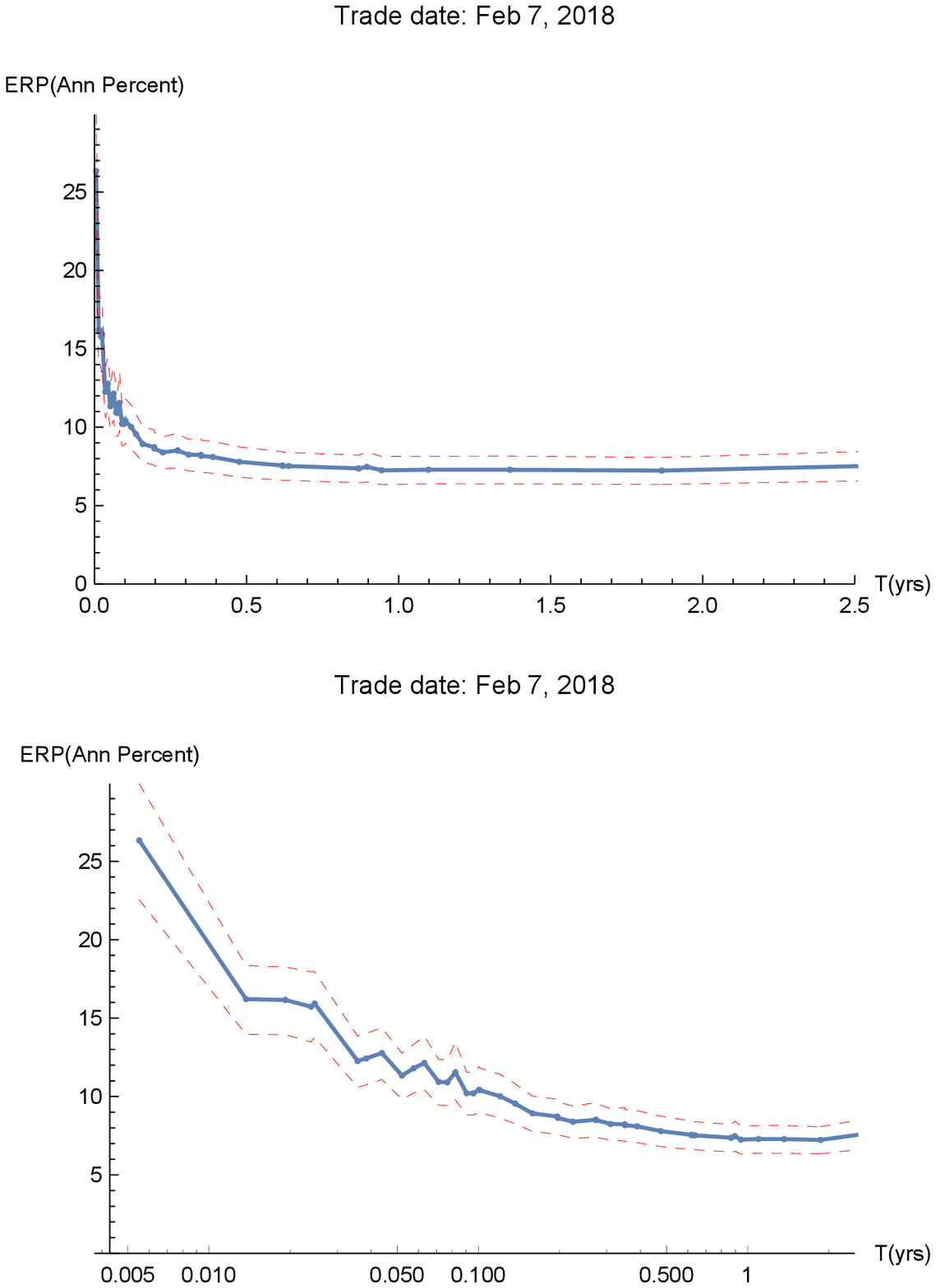}
  		\newline\newline
  	\end{center}		
  	\label{fig:ERPplot1}
  \end{figure}

\begin{figure}[tp] 
  	\caption{{\bf{Equity Risk Premium Term Structure: VIX at local low (10.85)}}
  	 \newline \hspace*{40pt} (Bands computed with CRRA $\kappa = 3 \pm 0.5$). }
  	\vspace{10pt}
  	\begin{center}
  		\includegraphics[width=0.8\textwidth]{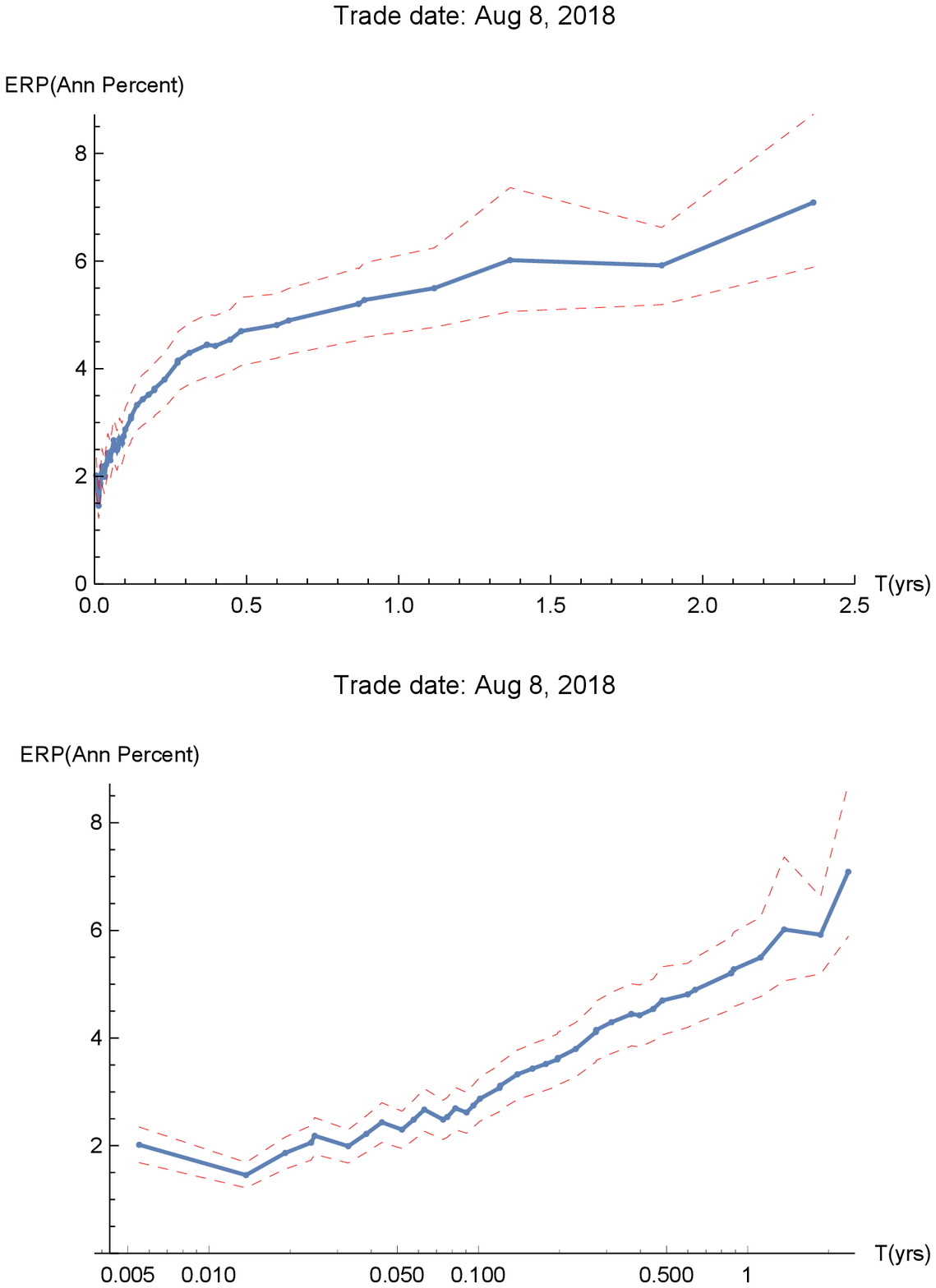}
  		\newline\newline
  	\end{center}		
  	\label{fig:ERPplot2}
  \end{figure}

\begin{figure}[tp] 
  	\caption{{\bf{Equity Risk Premium Term Structure: mid-teen VIX (16.21)}}
  	 \newline \hspace*{40pt} (Bands computed with CRRA $\kappa = 3 \pm 0.5$).}
  	\vspace{10pt}
  	\begin{center}
  		\includegraphics[width=0.8\textwidth]{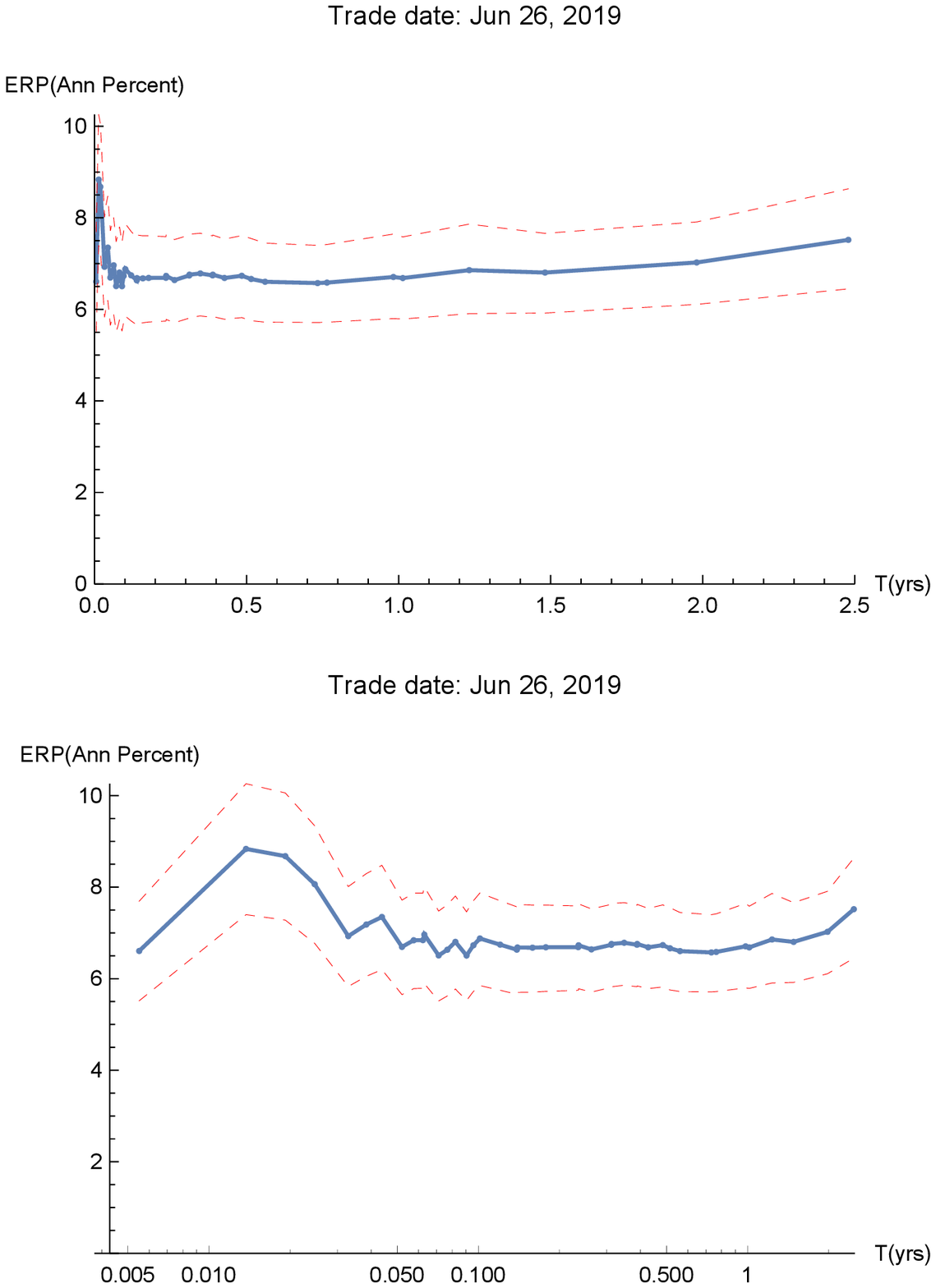}
  		\newline\newline
  	\end{center}		
  	\label{fig:ERPplot3}
  \end{figure}

\begin{figure}[tp] 
	\caption{\bf{Example of `Good' GMM fit: \newline
			Trade date: Feb 7, 2018. Expiration date: Feb 12, 2018}}
	\vspace{10pt}
	\begin{center}
		\includegraphics[width=0.8\textwidth]{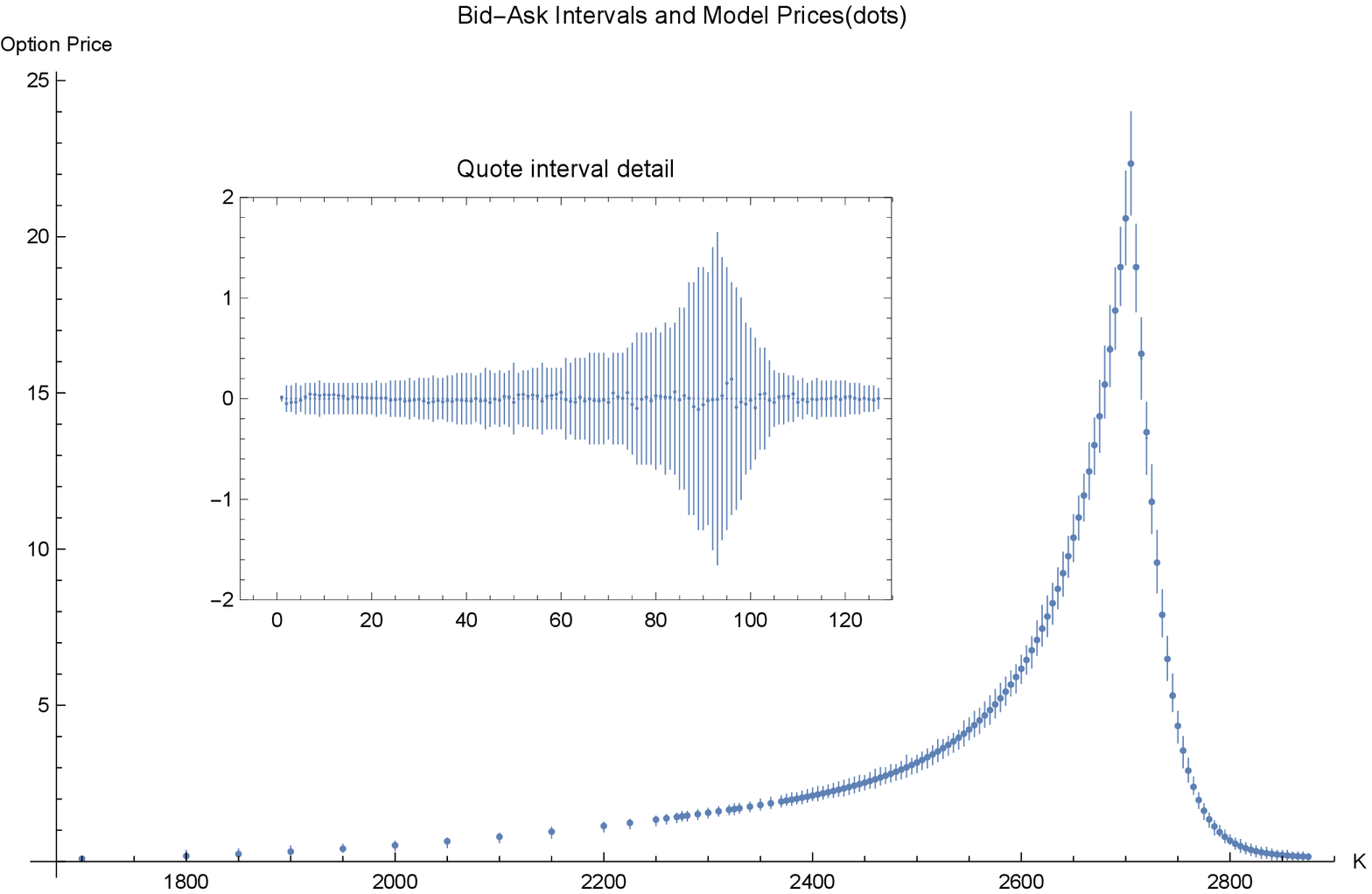}
	\end{center}		
	\label{fig:BidAskFitPlotG}
\end{figure}

\begin{figure}[b] 
	\caption{\bf{Example of `Acceptable' GMM fit: \newline
			Trade date: Feb 7, 2018. Expiration date: Feb 20, 2018}}
	\vspace{10pt}
	\begin{center}
		\includegraphics[width=0.8\textwidth]{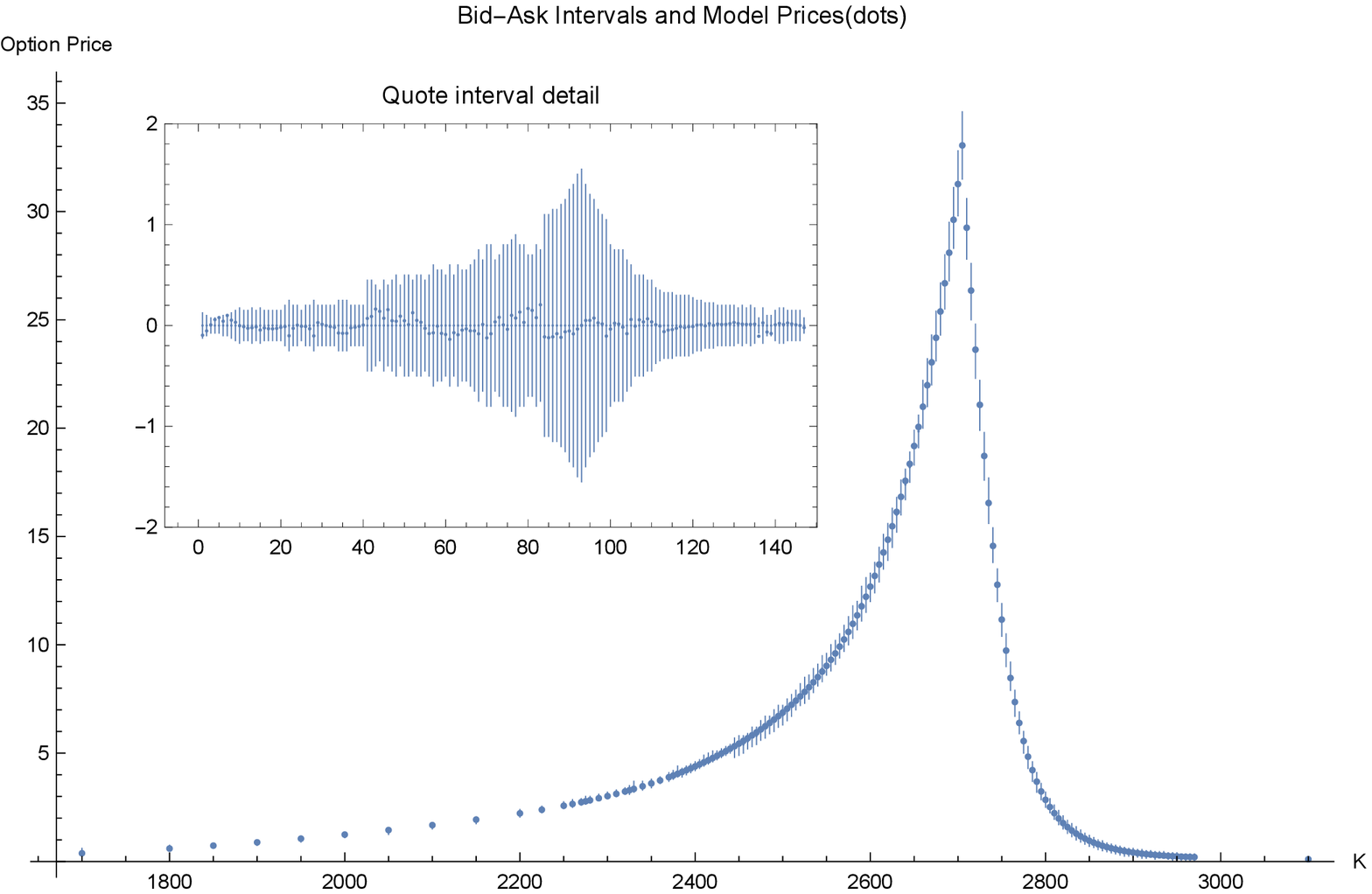}
	\end{center}		
	\label{fig:BidAskFitPlotA}
\end{figure}

\begin{figure}[tp] 
  	\caption{\bf{Smile and Extended Smile Fits: \newline
  			Trade date: Feb 7, 2018. Expiration date: Feb 12, 2018}}
  	\vspace{10pt}
  	\begin{center}
  		\includegraphics[width=0.8\textwidth]{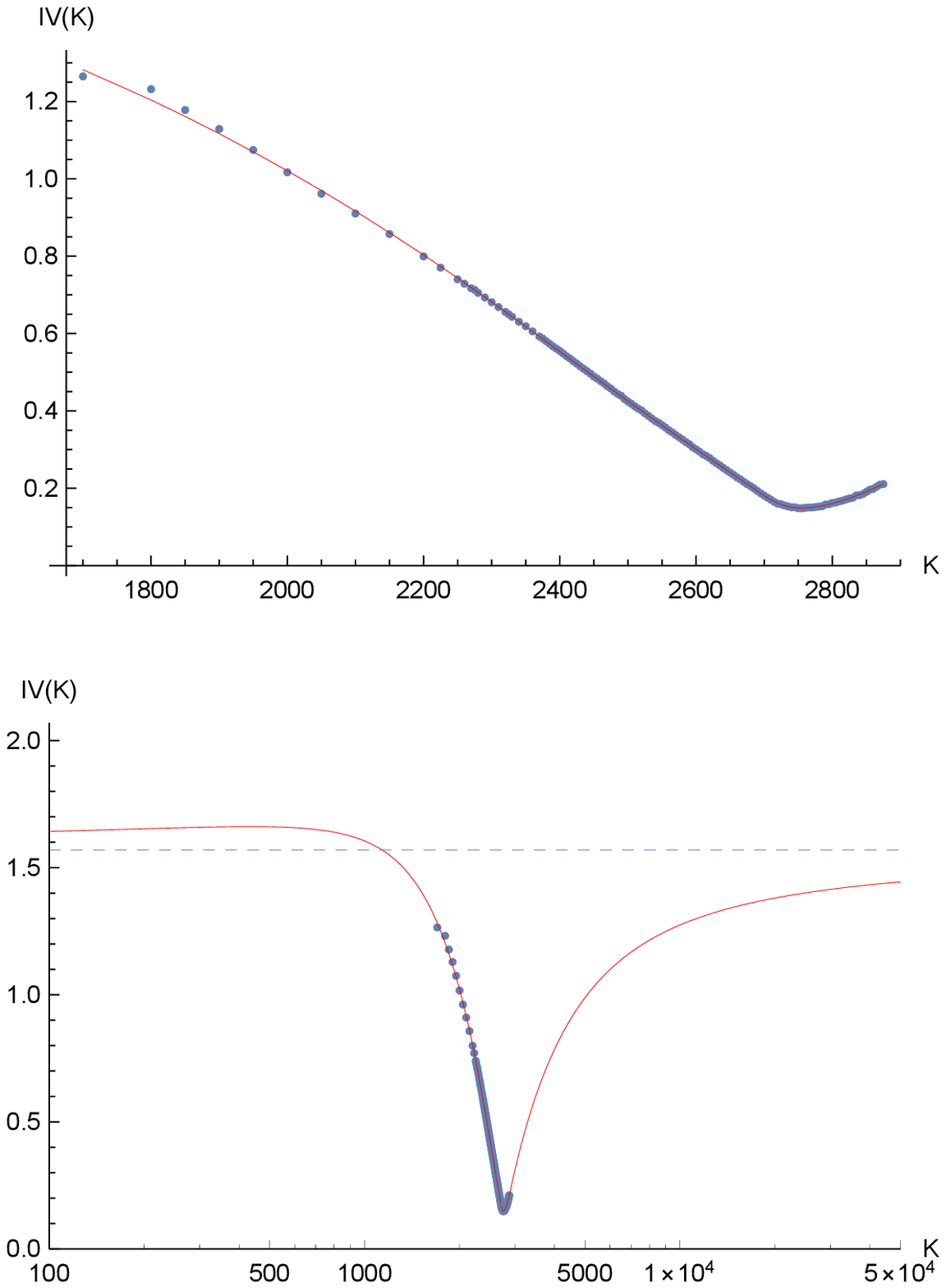}
  		\newline\newline
  	\end{center}		
  	\label{fig:SmilePlots}
  \end{figure}

    Figs. \ref{fig:ERPplot1}-\ref{fig:ERPplot3} show three examples of our new ERP term structure estimates.
    The top chart of each figure has a linear time scale. But, because the nearest expirations are so close together,
    it's hard to resolve the chart data. So, the bottom chart of each figure shows the same term structure with a
    logarithmic time scale. 
    
    Recall the trade dates were selected to have a relatively high, low, and middle-of-the-road volatility environments.
    In the higher volatility environment, the ERP estimates decay
    from a large value $\approx 26\%$ (from 2-day options) toward the longer-run averages (from 2.5 year options). 
    Conversely, in the lower volatility environment, the ERP term structure is increasing with the term, again toward unconditional values. Finally, the last volatility environment yields a relatively flat term structure.

    Qualitatively, those are quite plausible results. The key drivers here are volatility and risk-aversion. 
    However, anything that the market becomes concerned about, as long as it is reflected in the options market,
    can play a role in the results. 
    
    For Fig. \ref{fig:ERPplot1}, the ERP decay from 26\% is quite rapid, suggesting that
    the market viewed the Feb 5 volatility event as perhaps an ``internal technical event" -- 
    similar examples being the Oct 19, 1987 crash or May 6, 2010 Flash crash. In other words, the volatility
    jump was not associated with a systematic economic problem with long term persistence. 
    Instead, it was internal because the volatility increase -- way beyond what would be expected, given
    the SPX decline -- was exaggerated due to market internals. Internal factors on Feb 5
    included panic derivative trading under loss of liquidity and consequent disruption in the volatility product space -- events likely to have only temporary effects on the broader equity market.
    
    For Fig. \ref{fig:ERPplot2}, the `half-life' associated to the rising ERP looks longer to me than the
    half-life associated to the falling ERP in Fig.  \ref{fig:ERPplot1}.
    
     For Fig. \ref{fig:ERPplot3}, there is an interesting `hump shape' near term in an otherwise `flattish'
     term structure. I don't have any explanation for it. 
     
     For all the term structures, there are various small wiggles and oscillations. At this writing, my guess is that those are
     just natural data/estimation noise and not indicative of some fine structure in the market's `true' ERP's. 
     
     \subsubsection{Some literature compares}
     
     We have mentioned that \cite{duarte:2015} (henceforth D\&R) present several ERP term structure estimates by
     combining 20 models (their Fig. 4). They show results extending out to one month, one quarter, six month, one year, two years,
     and three years -- so 6 term points -- at various dates in their data set, which encompassed January 1960 to June 2103.
     In addition, they show an overall Mean term structure, which is very flat (rising from about 6\% to 7\%), and can be interpreted as their estimate of the unconditional ERP term structure.   They selected 3 dates which were peaks in the one-month-ahead
     ERP: Sept 1974, June 2012 and Dec 1982; and two dates which were low points in the same: Sept 1987, and Dec 1999.
     
     By my observation, Sept 1974 marked an S\&P500 index low in the 1973-75 recession, with an annualized volatility (using daily
     log-returns) of 30.5\%. The D\&R ERP term structure is generally downward sloping for that date, moving from
     about 15\%/year for the one-month horizon to about 8\%/year for the 3-year horizon. Given the elevated volatility, this
     is quite consistent with our results here. At the other extreme, Dec 1999 marked a relatively quiet month,
     with an annualized volatility of only 10.9\%. Note that this date was several months prior to the
     dot-com crash of Mar 2000-Oct 2002  and corresponding 2001-2002 recession.    
     For that date, the D\&R ERP term structure is generally upward sloping, moving from
     0 for the one-month horizon to about 6\%/year for the 3-year horizon. It is not monotone and does fall negative: to -1\%/year
     at the six month horizon -- something I have not yet seen in my method. 
     
     Overall, the approaches agree in that they both see significant time variation in the ERPs and a mixture of term structure shapes. I suspect they likely agree with ours in seeing an association of downward (resp. upward) sloping
     term structures with elevated (resp. below-average) volatility, although D\&R did not test for this association directly.
     The overall D\&R approach seems quite involved relative to ours, given its need to supply data for 20 diverse models. 
     We only need option prices (using the put-call parity regression method for cost-of-carrys). Finally, we note
     that our method is purely ``forward-looking", whereas the D\&R approach uses some regressions that
     are not predictive regressions.

   \newpage

   \section{Removing calendar spread arbitrage if you must} \label{sec:calarb}
     Recall that our optimizer in (\ref{eq:nominalobj}) is fitting the GMM model to
     \emph{mid-point quotes}; i.e., \newline $C^{mkt}_i = 0.5 \times (C^{bid}_i + C^{ask}_i)$, where $i$ indexes the strikes.
     If you could transact at mid-point quotes with no transaction costs (generally you can't), you
     would often find two types of arbitrage opportunities (`arb opp' for short):
     
     \begin{itemize}
     	\item butterfly spread arbitrage, and
     	\item calendar spread arbitrage.
     \end{itemize}
     The GMM fit, by construction, produces a model price free of butterfly spread arbitrage.
     The absence of butterfly arbitrage is equivalent to having a proper (non-negative)
     risk-neutral density $Q_{t,T}(S_T)$ and this occurs automatically with each fit.
     
     In contrast, nothing in our fitting procedure (so far discussed) prevents calendar spread
     arbitrage; indeed, you will see calendar spread arbitrage opportunities in fitted model prices.

     As a practical matter, the main violations are seen in the AM/PM \emph{dual expirations}, 
     referring to the Fridays where both an AM option (SPX) and PM option (SPXW) expire.
    For example, in Table \ref{tab:ERP020718}, there are seven such pairs. Using the file numbers of the first column,
    they are:
    
    \[ \{(4,5), \, (17,18), \, (22,23), \, (25,26), \, (28,29), \, (32,33), \, (35,36) \}.         \]
    In general, the criterion for calendar spread arbitrage involves the cost-of-carry parameters.\footnote{See \cite{fengler:2005}.}
     However, for the dual expirations, an arbitrage opportunity exists at $t<T_{AM}$ if:
    
    \be     \mbox{for some strike} \,\, K, \quad C_{t,T_{AM}}(K) > C_{t,T_{PM}}(K) \quad \mbox{or} 
           \quad    P_{t,T_{AM}}(K) > P_{t,T_{PM}}(K).   \label{eq:calarb} \eb 
    In words, if prior to expiration the (model) price of the AM expiration option exceeds the PM price, there is an arbitrage opp.
    This is a well-known criterion for  calendar spread arbitrage in an environment with zero dividends and interest.
    In our case, we do have dividends and interest. The reason (\ref{eq:calarb}) still applies
    is that exploiting the inequality involves a \emph{day trade}: opening and closing positions during the
    same regular trading session -- as we now show.\footnote{I adapt a nice discussion at \newline \url{https://quant.stackexchange.com/questions/15215/how-to-exploit-calendar-arbitrage}} 
    
    \pbold{Exploiting the arbitrage opportunity.} Let's review the trading under two assumptions:
    
    \begin{itemize}
    	\item[(A1)] no transaction costs or margin requirements.
    	\item[(A2)] dividends and interest accrue (or are owed) only to positions maintained \emph{overnight}. 
    \end{itemize} 
   Our assumptions combine aspects of `ideal' and `realistic' markets.    
    First, suppose at some $t < T_{AM}$, there is some strike $K$ such that $C_{t,T_{AM}}(K) > C_{t,T_{PM}}(K)$. 
    We sell the AM option and buy the PM option which generates an account credit 
    $x = C_{t,T_{AM}}(K) - C_{t,T_{PM}}(K) > 0$. At time $T_{AM}$, if $S_{T_{AM}} \le K$, the sold option expires
    worthless and the PM option can be sold for the non-negative amount $C_{T_{AM},T_{PM}}$. Thus, with no
    initial investment, we have earned a positive profit: $x + C_{T_{AM},T_{PM}} > 0$.   
    
    The other possibility is that $S_{T_{AM}} > K$, in which case the AM option is cash-settled
    at  $S_{T_{AM}} -  K$, a positive amount which we owe the option buyer. To meet that obligation, we
    borrow the stock and immediately sell it into the market at price $S_{T_{AM}} = (S_{T_{AM}} -  K) + K$.
    After paying the buyer, our account, at time $T_{AM}$, now consists of
    (i) cash, totaling $x + K$, and (ii) a short position in one share of the stock (index).
    Now, what happens at time $T_{PM}$? 
    
    If $S_{T_{PM}} > K$, we now receive $S_{T_{PM}} - K$ from the
    cash-settlement of the PM option, so our net cash is now $x + S_{T_{PM}}$. We immediately buy the
    stock in the market for $S_{T_{PM}}$, which closes our short position and leaves us with cash $x > 0$,
    our arbitrage profit.  
    
    On the other hand, if $S_{T_{PM}} \le K$, our long call expires worthless and we again buy the stock
    to close our short position for the price $S_{T_{PM}}$. This reduces our cash to  $x + K - S_{T_{PM}}$. 
    But  $x + K - S_{T_{PM}} \ge x > 0$ because $K \ge S_{T_{PM}}$. 
    
    Thus, with no initial investment, we are able to earn a positive profit of at least $x > 0$ under any
    eventuality, almost the classic definition of an arbitrage opportunity.\footnote{The classic definition is
    	weaker, requiring only a non-negative profit, strictly positive under some outcome.}  
    
    What is the effect of dividends and interest? Initially, we have
    options only and no dividends accrue (or are owed) to the account from $t$ through
    $T_{AM}$. On expiration day, the short sale of the stock would generate a potential obligation to pay dividends if the short position were maintained overnight, but it isn't: it's closed out the same day. That's the day trade part. 
    
    Interest earnings (or payments) can change the windfall profit from the exploit, but not its sign. For example, the initial credit $x$ may earn interest from $t$ to $T_{AM}$ in a positive interest rate environment.
    This would only increase our profit to some $x' > 0$. Conversely, although not currently a factor in the US, at this
    writing negative interest rates are common in Europe. With negative rates, that initial credit would be reduced 
    to some $x'$ by expiration day, where $0 < x' < x$. This is still positive -- so we still earn an arbitrage profit under (\ref{eq:calarb}).

    In summary, the cost-of-carry parameters $(r_{t,T},\delta_{t,T})$ can be neglected
    in determining if an arbitrage opportunity exists with dual expiration pairs:
    for those, criterion (\ref{eq:calarb}) suffices.

\begin{figure}[tp] 
  	\caption{{\bf{Example Removal of Calendar Spread Arbitrage}}. \newline
  			PM minus AM option prices. 
  			Trade date: Feb 7, 2018. Expiration date: Mar 16, 2018 \newline
  		Top: Market Data (dots) + original GMM fit to market mid-quotes (curve). \newline
  		Middle: Original GMM fit. \newline
  		Bottom: GMM fit to adjusted market quotes: Solid=PM adjusted. Dashed=AM adjusted.}
  	\vspace{10pt}
  	\begin{center}
  		\includegraphics[width=0.6\textwidth]{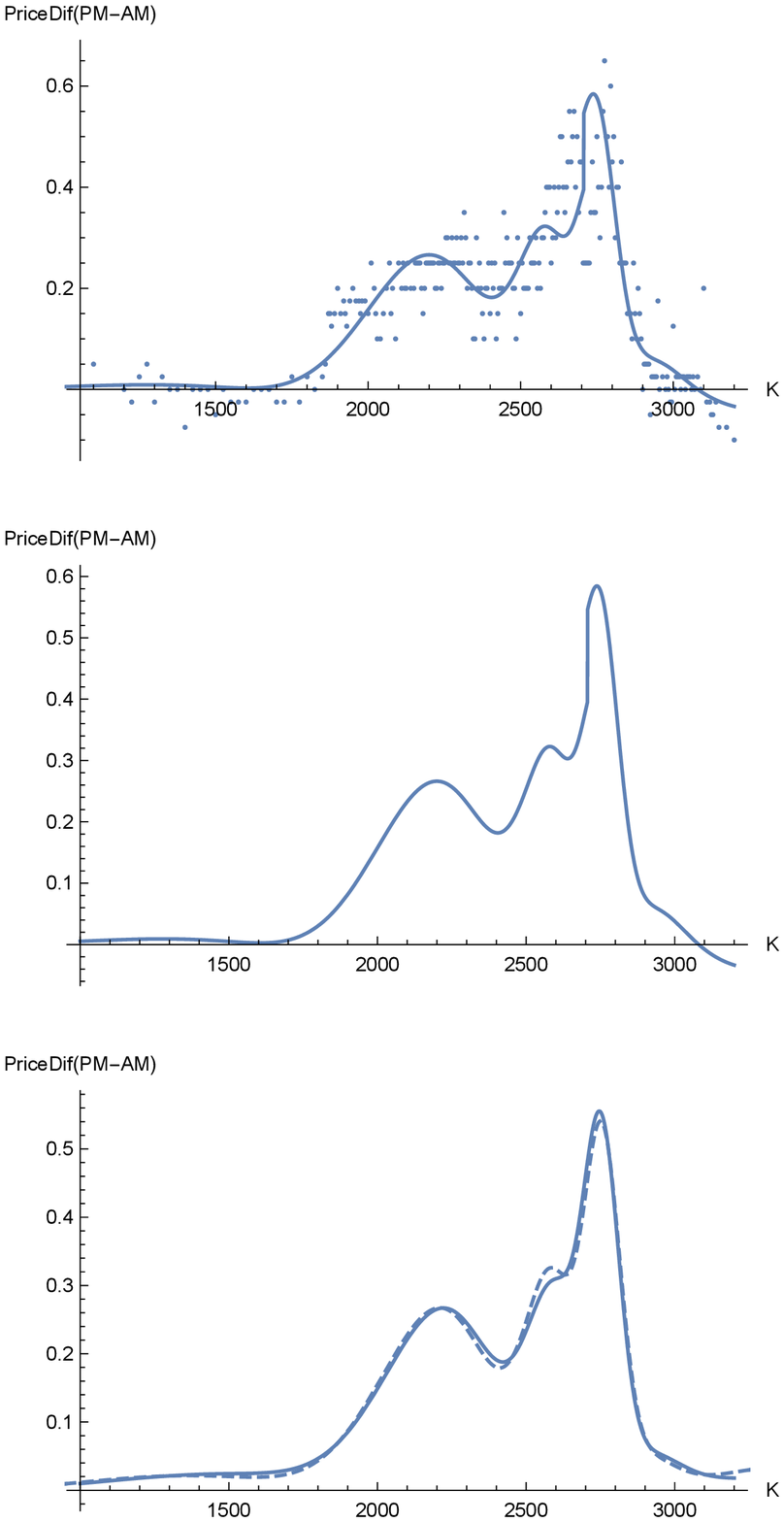}
  	\end{center}		
  	\label{fig:CalArbFix1718}
  \end{figure}

\begin{figure}[b] 
  		\caption{{\bf{Example Removal of Calendar Spread Arbitrage}}. \newline
  		PM minus AM option prices. 
  		Trade date: Feb 7, 2018. Expiration date: Dec 21, 2018 \newline
  		Top: Market Data (dots) + original GMM fit to market mid-quotes (curve). \newline
  		Middle: Original GMM fit. \newline 
  		Bottom: GMM fit to adjusted market quotes: Solid=PM adjusted. Dashed=AM adjusted.} 
  	\vspace{10pt}
  	\begin{center}
  		\includegraphics[width=0.6\textwidth]{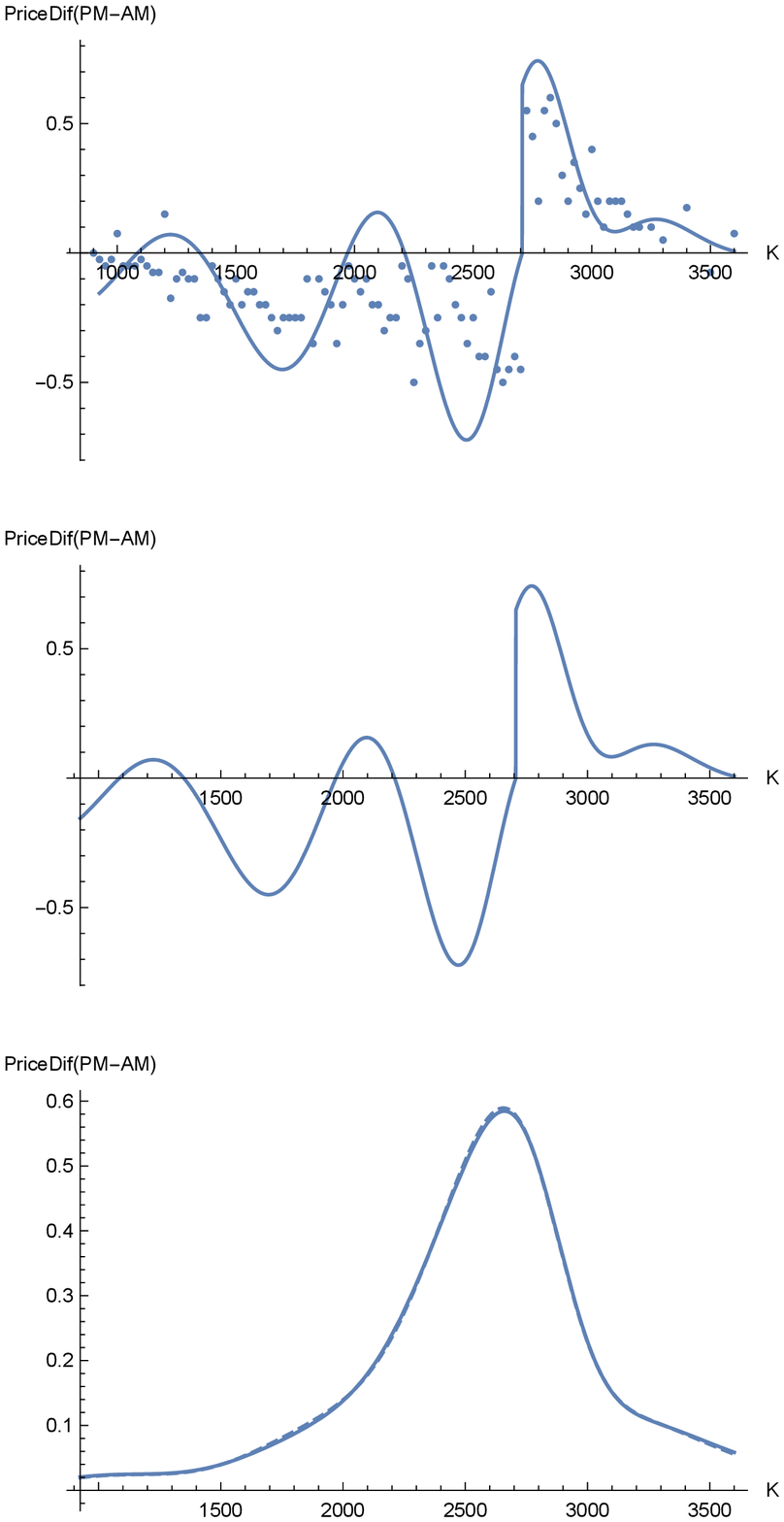}
  	\end{center}		
  	\label{fig:CalArbFix3536}
  \end{figure}

\begin{figure}[tp] 
  	\caption{\bf{RNDs (Q) and RWDs (P): Feb 7, 2018}}
  	\vspace{10pt}
  	\begin{center}
  		\includegraphics[width=0.7\textwidth]{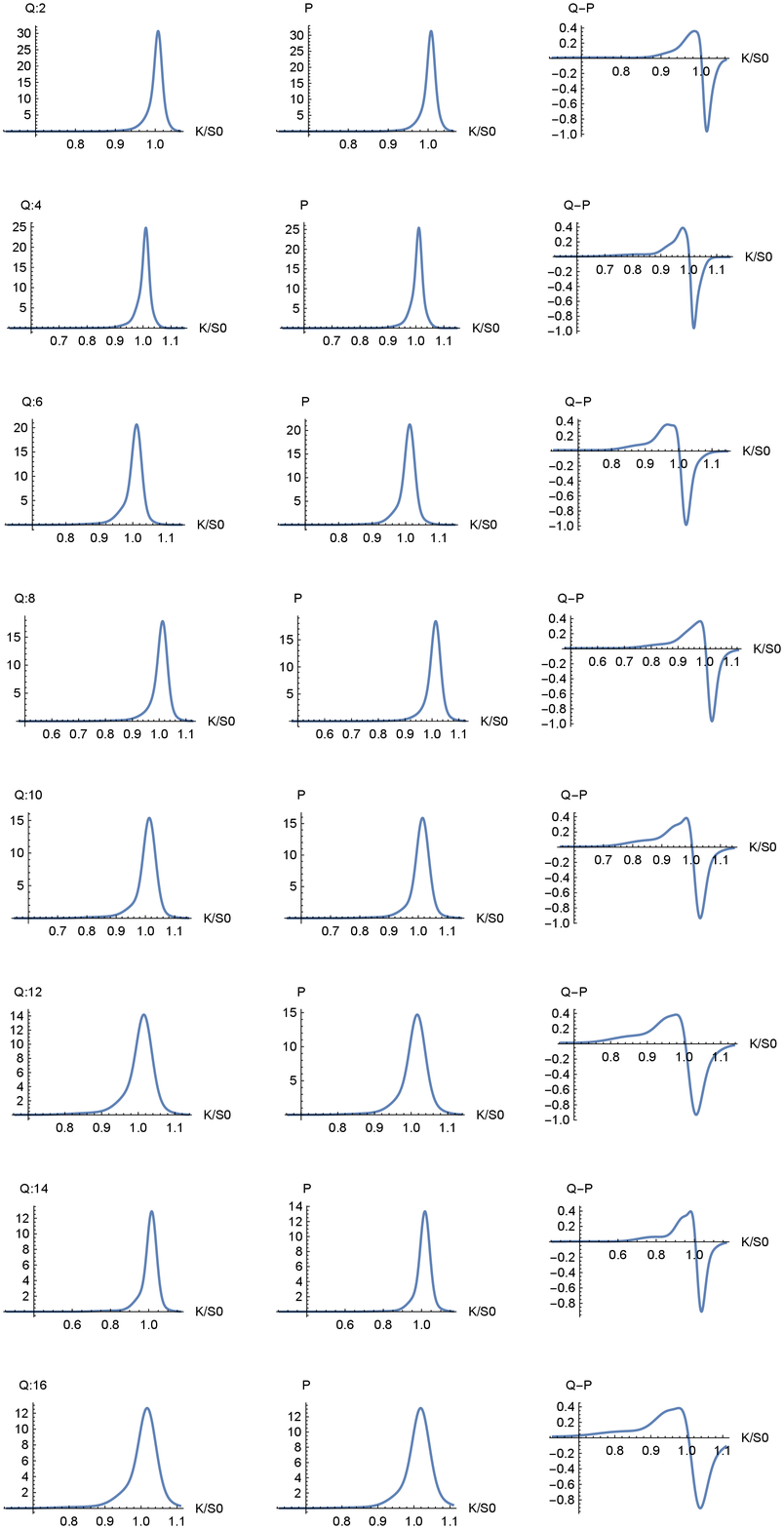}
  		\newline\newline
  	\end{center}		
  	\label{fig:pdfsA}
  \end{figure}

\begin{figure}[tp] 
	\caption{\bf{RNDs (Q) and RWDs (P): Feb 7, 2018 (cont.)}}
	\vspace{10pt}
	\begin{center}
		\includegraphics[width=0.7\textwidth]{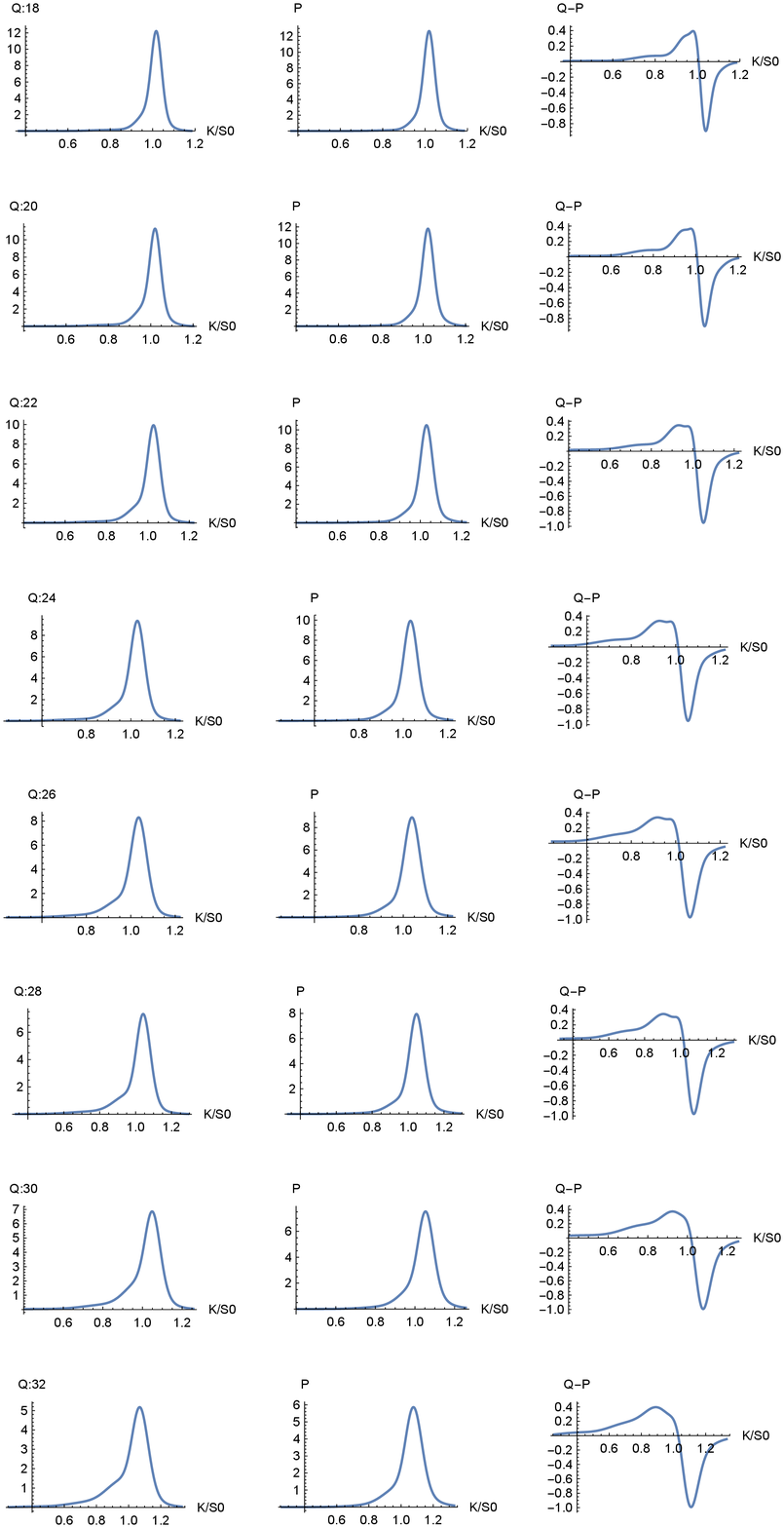}
		\newline\newline
	\end{center}		
	\label{fig:pdfsB}
\end{figure}

\begin{figure}[tp] 
  	\caption{\bf{RNDs (Q) and RWDs (P): Feb 7, 2018 (cont.)}}
  	\vspace{10pt}
  	\begin{center}
  		\includegraphics[width=0.7\textwidth]{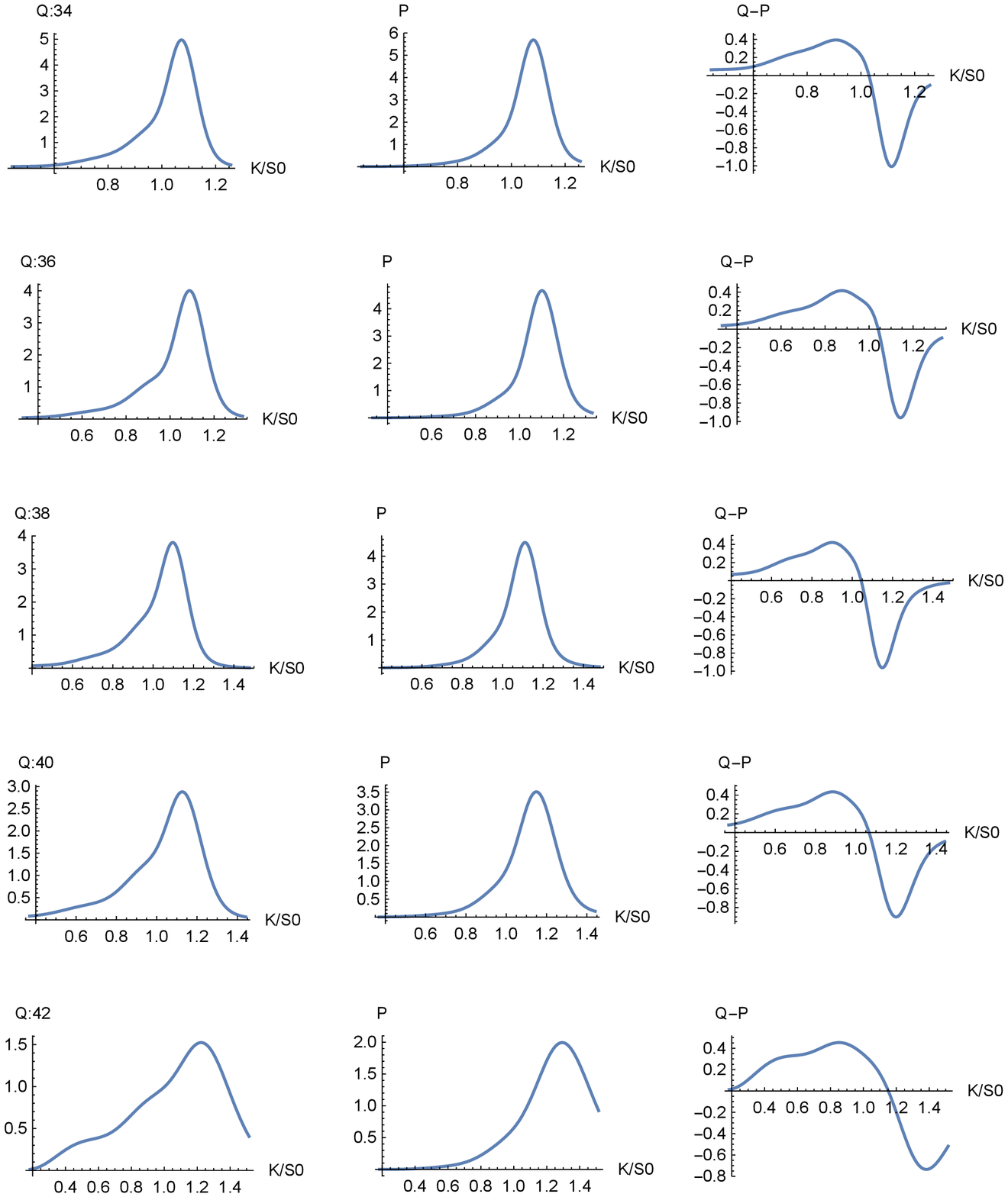}
  		\newline\newline
  	\end{center}		
  	\label{fig:pdfsC}
  \end{figure}

    \pbold{Calendar arbitrage in GMM fits and its removal.} It's quite common to see (\ref{eq:calarb}) hold 
    (for some ranges of strikes) in the GMM fits to our data. For two examples, consider the two expiration pairs (17,18) and (35,36) from Table \ref{tab:ERP020718}. Figures \ref{fig:CalArbFix1718}-\ref{fig:CalArbFix3536}
    show the situation. There are three charts per figure. The top chart shows the difference between
    the PM and AM option prices both in the data (the dots) and the GMM fit (the curve). The middle chart shows
    just the GMM fit differences for clarity. The bottom chart shows the `fix' -- explained below. Recall we always use
    out-of-the-money options, so the difference is $P_{t,T_{PM}}(K) - P_{t,T_{AM}}(K)$ for $K < S_t$ and 
     $C_{t,T_{PM}}(K) - C_{t,T_{AM}}(K)$ for $K \ge S_t$. For the (17,18) expiration, there are 21 strikes (dots) with
     negative values; for the (35,36) expiration there are 71.
     
     What causes calendar arbitrage in the model? One cause is calendar arbitrage in the \emph{data}. In Fig. 
     \ref{fig:CalArbFix1718}, you can see that  $C^{GMM}_{t,T_{PM}}(K) < C^{GMM}_{t,T_{AM}}(K)$ for strikes
     above $K = 3100$. If you look at the market data (the dots), although quite noisy, it's clear that
     the GMM fit is roughly following the data, which is a good thing. The data turns negative above $K = 3000$,
     and the model fit is following it down. The same type of thing is seen in the top chart of Fig. \ref{fig:CalArbFix3536}.   
     
     The main ingredient to my `fix', which is shown in the bottom chart of the figures is simply
     to shift the market price that is being fitted by a small amount, so that the calendar arbitrage
     is no longer present in the data. Recall that we are fitting to the mid-point quote, which
     is the average of the bid and ask quote. Since our secondary objective is to get a model
     fit between the bid and ask, fitting to the mid-quote is the natural choice. But there is nothing
     sacrosanct about the mid-quote. We can fit to some other price within the bid-ask range if it is
     convenient. Which prices should be shifted? 
     
     First, I shift the market price of the PM option by a
     small positive increment when there was a calendar arb violation at the mid-quote vs. the corresponding
     AM option. The AM option price is left unaltered. Specifically, for the examples shown, I chose the increment to be $0.005 \,\, \times$ the 
     mid-quote price ($\smallfrac{1}{2}$ of 1\%) or $\$0.01$, whichever was larger. With that rule (and certain other
     data `cleanups' explained below), the resulting GMM fits are
     shown as the solid curves in the bottom charts of  Figures \ref{fig:CalArbFix1718}-\ref{fig:CalArbFix3536}. 
     
     The other way to go about it, which I do next, is to \emph{decrease} the market price of the AM option by the
     same rule, and re-run the fits. In this case, the PM option prices is left unaltered. 
     Those results are plotted as the dashed curves in the bottom charts of Figures \ref{fig:CalArbFix1718}-\ref{fig:CalArbFix3536}. The two curves are very close, so you have to look carefully to see the
     dashing. As you can see, the fits are now calendar-arb-free under either adjustment procedure.  
     
     Before explaining the other data clean-up issues, let me show the effect of the adjustments on the
     corresponding row entries in Table	\ref{tab:ERP020718}. In Table \ref{tab:BeforeAfterCalArbAdj}, the
     top panel repeats the previous results from Table \ref{tab:ERP020718}. The middle panel shows
     the results of the fit after adjusting (upward) the market prices for the PM options. As you can see, the ERP's have increased somewhat for the PM options (SPXW). This is not surprising, as higher option prices means higher implied volatilities,
     which suggests higher ERP's for a risk-averse agent. Finally, the bottom panel shows
     the results of the fit after adjusting (downward) the market prices for the AM options. Those ERP's have 
     decreased, again not surprising by the same rationale.

     \pbold{Other data cleanups/adjustments as part of calendar arb removal.} 
     Recall the index forward price at trade date $t$ for option expiration
     $T$ is $F_{t,T} = S_t \exp\{(r_{t,T} - \delta_{t,T}) (T-t)\}$. With the VIX white paper method for the
     cost-of-carry parameters (under the original procedures), I took the risk-free rate $r_{t,T}$ from the US Treasury (interpolated) yield curve, using the \emph{whole number} of calendar days from $t$ to $T$. This meant that, for paired expirations, the
     AM and PM values of $r$ were identical. However, since $T-t$ differed for the AM and PM expirations,
     the factors $\exp\{(r_{t,T} (T-t)\}$ were slightly different. Also, the AM and PM option-implied
     forward prices were slightly different, which meant that the dividend yields $\delta$ were
     slightly different. For example, for the (17,18) paired expiration
     $(r,\delta,F)_{AM} = (0.01376,0.02376,2703.76)$ and  $(r,\delta,F)_{PM} = (0.01376,0.02314,2703.91)$.
     
     Now, the day trade discussion above suggested to me that
     a better setup would be to enforce \emph{identical forward prices} for the AM and PM expirations. That way,
     the implied dollar dividends and implied dollar interest earnings would be the same for the two expirations.
     That notion was used in the discussion above, and corresponds to the market practice that interest and dividend
     earnings are based upon \emph{overnight} holding periods. Since the number of overnight holding periods for
     $(t,T_{AM})$ and $(t,T_{PM})$ are identical, so should be the corresponding forward prices. So, one additional
     data adjustment here takes the PM option-implied forward and corresponding $(r,\delta)$ as ``correct" -- since those are based upon the more standard close rather than open. Then, adjust $(r_{AM},\delta_{AM})$ so that
     $r_{AM} T_{AM} = r_{PM} T_{PM}$ and    $\delta_{AM} T_{AM} = \delta_{PM} T_{PM}$. For the (17) expiration, the
     cost-of-carrys after this adjustment was  $(r,\delta,F)_{AM} = (0.01387,0.02331,2703.91)$. This is one reason
     why you see slightly different results (compared to the top panel) for the `After' fits in Table \ref{tab:BeforeAfterCalArbAdj} for \emph{both} AM and PM fits.
     
     You will also notice from Table \ref{tab:BeforeAfterCalArbAdj} that, for the (35,36) expiration, the adjustment included a different PutBidMin cutoff and consequent different number of options fitted. The reason for that adjustment was that, with PutBidMin=\$0.05, the (35) expiration had option quotes extending down to $K=200$ with a corresponding mid-quote IV of 0.85. But the (36)
     expiration had option quotes extending down only to $K=900$ with a mid-quote IV of 0.45. Now the GMM fit tends
     to produce a $\sigma_{max}$, the maximum of the fitted volatilities $\{\sigma_i\}$, that roughly moves up and
     down with the maximum IV of the data. With such a large disparity in the max IV's of the data, one would
     expect to find $\sigma_{max}(AM) > \sigma_{max}(PM)$ and indeed that was the result: $\sigma_{max}(AM) = 0.65$ and
     $\sigma_{max}(PM) = 0.47$. That discrepancy will lead to 
     asymptotic calendar arb violations as $K \ra 0$ or $K \ra \infty$.\footnote{Equivalent to (\ref{eq:calarb})
     	under our assumptions is $IV^2_{AM}(K) \, T_{AM} >  IV^2_{PM}(K) \, T_{PM}$ for a calendar arb violation. As  $K \ra 0$ or $K \ra \infty$, that translates to $\sigma^2_{max}(AM) T_{AM} >  \sigma^2_{max}(PM) T_{PM}$. But since $T_{AM}$ and $T_{PM}$ are
     so close, $\sigma_{max}(AM) = 0.65$ and $\sigma_{max}(PM) = 0.47$ ensures a violation.}        
     To avoid those, I boosted PutBidMin to \$0.15, which
     resulted in more aligned data with a minimum strike of $K=925$ for both AM and PM, and consequent
     adjusted fit $\sigma_{max}(AM) = 0.46$ and $\sigma_{max}(PM) = 0.47$, resolving that issue. In response,
     the ERP for the (35) expiration moved down slightly.
     
     \newpage

      \begin{table}[th] 
     	\caption{\bf{Feb 7, 2018: ERP before and after calendar spread arbitrage adjustments.}} 
     	\begin{center}
     		\begin{tabular}{llcc}
     		\midrule 
     	File/ & Expiration       & \multicolumn{2}{c}{ERP}  \\
     	Root & Date (days-to-go) & Before & After \\		
     	\cmidrule(r){1-1}   \cmidrule(r){2-2} \cmidrule(r){3-4}      
     		17 SPX&Mar 16, 2018 (37)    &  10.43  & 10.31 \\
     		18 SPXW&Mar 16, 2018 (37)   &  10.39  & 10.49 \\
     		35 SPX&Dec 21, 2018 (317)   &   7.37  & 7.335 \\
     		36 SPXW&Dec 21, 2108 (317)  &   7.36  & 7.375 \\ 		
     		\midrule 
     \end{tabular}
 \end{center}
\label{tab:BeforeAfterSummary}
\end{table}	

   \pbold{Final adjusted estimates.}   
Our final adjusted estimate for the ERP's is the simple average of the ERP from Table \ref{tab:BeforeAfterCalArbAdj}
under the two adjustment procedures. Those final estimates are shown in Table \ref{tab:BeforeAfterSummary}.
For example, the first row of Table \ref{tab:BeforeAfterSummary} has an `After' value of 10.31, which the average
of 10.42 from the first row of Table \ref{tab:BeforeAfterCalArbAdj}/Panel II and   
10.20 from the first row of Table \ref{tab:BeforeAfterCalArbAdj}/Panel III. 	
     
     \pbold{Are these adjustments worth the trouble?} If one were trying to generate an arbitrage-free implied
     volatility surface from GMM fits, then calendar arb violations should be removed. 
     
     However, we aren't doing that --
     instead, we are trying to estimate ERP term structures. As you can see from Table \ref{tab:BeforeAfterSummary}, the ERP
     changes due to these adjustments (at least for these examples) are relatively modest. 
     
     One issue is whether
     or not you feel the need to distinguish the AM ERP's from the PM ERP's. We have done so here.
     But, if you don't, you can just average
     the original (unadjusted) pair ERP values and be done with it. For example, for Table \ref{tab:BeforeAfterSummary}, those
     pair averages are, for the Mar 16 expiration: 10.41 (before adj.) and 10.40 (after adj.). Also, for
     the Dec 21 expiration: 7.365 (before adj.) and 7.355 (after adj.) -- clearly, those tiny differences are not worth the trouble. 
    
     If you do want to distinguish the AM/PM ERP's, you can attempt some adjustment as we have done and then decide if the differences matter for your purpose. As a final caveat, we remind the reader of Footnote \ref{ft:bustime}, which mentions the
     possibility of needing business time vs. calendar time corrections for very short-dated dual expirations.

 \section{Sensitivity of the ERP to the cost of carry methodology} \label{sec:altcoc}
   The ERP results shown in Figs. \ref{fig:ERPplot1}-\ref{fig:ERPplot3} and corresponding
   Tables \ref{tab:ERP020718}-\ref{tab:ERP080818} use the `VIX white paper method' for $(r,\delta)$'s --
   a method explained in Sec.  \ref{sec:VIXwp}.
   
   For that method, the risk-free rate is the corresponding US Treasury rate for the same maturity as the
   option expiration. Is that the best choice? The average effective financing (or investing)
   rate for professional option traders and market makers is not directly observable. Take financing, for example. Typically,
   a market making firm will have (perhaps several) so-called ``prime brokers", which provide custody,
   clearing, securities lending, financing, and other services. As the prime broker is typically 
   a division of a large investment/money center bank, the bank/broker  is able to borrow dollars at benchmark rates like
   Fed Funds and 
   LIBOR (overnight and term).\footnote{While USD LIBOR is currently referenced in financial contracts with some \$200 trillion in notional value, it's likely going to be replaced as a reference rate as early as 2021. U.S. monetary authorities have identified the Secured Overnight Financing Rate (SOFR) as the rate that represents best practice for use in certain new USD derivatives and other financial contracts. See \url{https://www.newyorkfed.org/arrc/sofr-transition} for more about the planned transition from LIBOR.}   
   	
   	 Consequently the market maker will be offered financing at benchmark rates plus a (non-observable) negotiable
   spread. Another non-observable is the term: market makers may choose to finance their book at overnight rates or perhaps lock-in a term rate. Nevertheless, these considerations do suggest financing rates
   closer to overnight LIBOR plus some spread. Excess cash might be swept into an institutional money
   fund. The net result is we have an effective risk-free rate $r$ -- not observable but likely different 
   than the term US Treasury rate. 
   
   Additional issues occur with the dividend yields we have estimated.  To take an example,
   consider Expiration (File number) 31 in Table \ref{tab:ERP020718}, which is the Feb 7, 2018 trade date with
   option expiration in 174 days, so almost half a year out. If you look at the corresponding row entry
   in Table \ref{tab:ERP020718CoC} you'll see the VIX white paper method produced
   an $r = 1.70\%$ and $\delta = 1.34\%$. From the S\&P Dow Jones Indices website\footnote{\label{ft:spindices}\url{https://us.spindices.com/indices/equity/sp-500-dividend-points-index-annual}},
   the total dividend points earned from a holder of the S\&P500 index in 2017 was $D = 49.01$. This means the
   trailing 12-month (TTM) dividend yield on Feb 7, 2018 was approximately $\delta_{TTM} \approx 49.01/2706.48 = 1.81\%$.
   So, if you simply projected the historical dividend yield forward, it makes the VIX white paper $\delta$ look low.
   Indeed, the subsequent \emph{realized} yield over this 174 day period was actually $\delta = 1.94\%$ (see below). 
   Also, a glance at FRED\footnote{\url{https://fred.stlouisfed.org/series/USD6MTD156N}} shows LIBOR on that date was 
   $r_{LIBOR} = 1.44\%$ (Overnight), while $r_{LIBOR} = 2.00\%$ (6-month).   
   
   Because of these various ambiguities, I investigated a second cost of carry method.
    
  \subsection{Cost of carry: a put-call-parity regression method.} The method is very easy. 
  Fixing a trade date $t$ and expiration $T$, and indexing the available strikes as $K_i$, we can 
  write the put-call parity relation (\ref{eq:putcallparity}) as
  
  \[   \left(\frac{P_i - C_i}{K_i} \right) = e^{-r \tau} -  \e^{-\delta \tau} \left(\frac{S_t}{K_i} \right). \]   
  This suggests simply estimating the OLS regression
  \be y_i = a + b x_i + \epsilon_i,  \label{eq:PCparitymethod} \eb 
  where $y_i = (P_i - C_i)/K_i$ and $x_i = S_t/K_i$, where $(P_i,C_i)$ are the mid-quotes for strike $K_i$.
  The estimated coefficient $(a,b) = (e^{-r \tau},-e^{-\delta \tau} )$ and $(r,\delta)$ estimates follow.
  Let  $(r_{reg},\delta_{reg})$ denote `regression' estimates. 
  They are seen in the right half of  Table \ref{tab:ERP020718CoC}.
  
  In the last table column, I show the subsequent `Realized yield', call it $\delta_{real}$, known (at this writing) for 
  all expirations except the last two. It was computed from the relation $\delta_{real} \times T = \log(1 + \mbox{DIVPTS}/S_0)$,
  where $T = \mbox{days}/365$ (whole number of days). Here DIVPTS is the S\&P500 dividend points earned between the two dates,
  computable from data provided at the website in Footnote \ref{ft:spindices}. 
  As one sees, for Expiration 31, we now find $r_{reg} = 2.36\%$, $\delta_{reg} = 1.97\%$, and $\delta_{real} = 1.94\%$.
  So,  $\delta_{reg}$ was a good forecast. Note that the forward price estimates are quite
  close between the methods -- thus, increases in $\delta$ with the regression method imply increases in $r$ to keep the forward price about the same. 
  
  Unfortunately, a comparison of all expirations shows the regression $\delta's$
  are an evident improvement only for the more distant half of the expirations, say Expiration 21 and beyond. For expirations less than
  60 days out, neither method seems to be a particularly good forecaster of realized dividends.
 
 For each option expiration regression, I used every available strike: (i) with a put-call pair, and (ii) 
 with non-zero bids.
  If you plot the data in (\ref{eq:PCparitymethod}), it's virtually indistinguishable from the fitted line, with the various
  $R^2$ (and adjusted $R^2$) all extremely close to 1. Specifically, I found $1 - R^2$ in the range
  $10^{-8}$ to $10^{-6}$ for all expirations. 
  
  The key issue is: what happens to the ERP's? The ERP columns in Table \ref{tab:ERP020718CoC} are highlighted in bold. They show generally a difference less than 0.1 percentage points between the two methods. That is, $|\mbox{ERP}_1 - \mbox{ERP}_2| < 0.1$, when the ERP's are reported in annualized percents as I do.

  The story for the Aug 8, 2018 trade date is similar. For that one, in Table \ref{tab:ERP080818CoC}, I just show
  the first and last 4 files for brevity. Again, while the $(r,\delta)$'s change significantly under the two methods,
  the option-implied forward prices are close, and this leads to very similar ERP's. 
  
  My conclusion is that either cost-of-carry method is suitable for the ERP: use whatever is convenient.
  The cost-of-carrys from the regression method are not uniformly an improvement over the VIX white paper method -- 
  although they do seem closer to the experienced costs for expirations at least 2 months distant. If all you seek is an ERP estimate,
  these $(r,\delta)$ differences may not matter.

  \section{The order of the density transformations}
  
  Let $\mathcal{T}_{\kappa}$ denote an exponential change-of-measure transformation with parameter $\kappa$.
  That is, $\mathcal{T}_{\kappa}$ transforms suitable probability densities $f(x)$ into new probability densities. 
  ($f$ is suitable if $\e^{-\kappa x} f(x) \in L^1$). More specifically, define
  $f^{(\kappa)} = \mathcal{T}_{\kappa}(f)$ to mean:
  
  \be  f^{(\kappa)}(x) = \mathcal{T}_{\kappa}(f)(x) \equiv 
            \frac{\e^{-\kappa x} \, f(x)}{\int \e^{-\kappa x}  f(x) \, dx}.   \label{eq:Tkappa} \eb 
  With that notation, the transformations of (\ref{eq:pqduality}) can be written succinctly as
  \be    q_{\bar{X}_T} =    \mathcal{T}_{\kappa}(p_{\bar{X}_T}) 
            \quad \Leftrightarrow \quad  p_{\bar{X}_T} =    \mathcal{T}_{-\kappa}(q_{\bar{X}_T}).  \label{pqdualformal} \eb     
  Recall that these are transformations of log-total-return pdf's, hence the `bars'. From options, we find RND's
  for the log-\underline{price}-returns; we needed to move from those to log-total-returns because the latter is
  what follows from the market-clearing equilibrium model of the Appendix. That move was accomplished by
  another simple transformation, $\mathcal{S}_c$, a shift operator. Here, if $f(x)$ is \emph{any} pdf on the real line,
  and $c$ is a real number, then $g = \mathcal{S}_c f$  means $g(x) = f(x-c)$. Obviously $g(x)$ is also a pdf.
  With that notation, we start with $q_{X_T}(x)$ from the GMM option fit, and get $q_{\bar{X}_T}(x)$ from
  $q_{\bar{X}_T} =  \mathcal{S}_c \, q_{X_T}$, where $c = \delta_{t,T} \times (T-t)$; this is simply (\ref{eq:divadj)}) again. 
  Combining,  
  \be        p_{\bar{X}_T} =    (\mathcal{T}_{-\kappa} \,  \mathcal{S}_c) (q_{X_T}), \quad \mbox{where} \quad 
         c = \delta_{t,T} \times (T-t).  \label{eq:combinedtransf} \eb 
  A natural question: could we have reversed the order? That is, could we have started with
  the risk-neutral log-price-return density, next performed an exponential change of measure, then shifted to a
  log-total-return density? It's easy to see the answer is `yes'. After all, if $f$ is an arbitrary (suitable) pdf,
  and $c$ is any real number, then
  
  \[ g_1(x) \equiv (\mathcal{T}_{\kappa} \,  \mathcal{S}_c) (f)(x) =
              \frac{\e^{-\kappa x} \, f(x-c)}{\int \e^{-\kappa x}  f(x-c) \, dx} =
               \frac{\e^{-\kappa (x-c)} \, f(x-c)}{\int \e^{-\kappa x}  f(x) \, dx}, \]
               while
   \[ g_2(x) \equiv ( \mathcal{S}_c \, \mathcal{T}_{\kappa}) (f)(x) =
    \mathcal{S}_c \frac{\e^{-\kappa x} \, f(x)}{\int \e^{-\kappa x}  f(x) \, dx} =
   \frac{\e^{-\kappa (x-c)} \, f(x-c)}{\int \e^{-\kappa x}  f(x) \, dx}. \]
   So, $g_1 = g_2$ and the operations \emph{commute}. There are two distinct, but equivalent, ways to 
   move from the source, $q_{X_T}$, to the target: the real-world log-total-return density, $p_{\bar{X}_T}$.
   Schematically, with the source density at the upper left and the target density at the lower right:

     \[\begin{tikzcd}
   q_{X} \arrow{r}{\mathcal{S}_c} \arrow[swap]{d}{\mathcal{T}_{-\kappa}} & q_{\bar{X}} \arrow{d}{\mathcal{T}_{-\kappa}} \\
   p_{X} \arrow{r}{\mathcal{S}_c} & p_{\bar{X}}
   \end{tikzcd}
   \]

    \newpage
  
  \section{Conclusions}
  
  We've explored in detail an attractive method to estimate  ERP term structure. Risk-neutral densities
  are estimated by fitting Gaussian mixture models to option quotes. Real-world densities follow via an
  exponential change of measure. Finally, ERP's are found by a simple analytic formula, (\ref{eq:ERPfinal}), 
  using fitted parameters plus $\kappa$.  We used $\kappa = 3 \pm 0.5$, a 
  plausible range based upon the historical record. In any event, under our assumptions,  (\ref{eq:ERPfinal}) holds for arbitrary $\kappa$.  
  
  \Pbreak
  Examples used relatively high, low, and middling volatility environments. The estimated ERP term structures  
  seemed sensible and support earlier research concluding ERP term structures have significant time variation.
  While generally not monotone with the term, examples were roughly downward sloping, upward sloping, or flattish.
  As one would hope, the example term structures tended toward unconditional ERP estimates at large maturity. 
  Moreover, our estimates were shown relatively insensitive to both (i) changes in the cost-of-carry method and (ii) the removal of small calendar-arbitrage violations.

  Calendar arb violations were driven largely by violations in the data,
  specifically in the mid-quotes at the dual SPX/SPXW expirations; we gave a method to remove them from the GMM fits, and consequent
  ERP estimates, if desired. If you don't feel compelled to distinguish AM/PM ERP's, which represent horizons only 6\smallfrac{1}{2}
  hours apart, I'd say it's a non-issue.
  
  \Pbreak
  Computations were done in Mathematica and were sometimes tedious. (Table \ref{tab:ERP020718}-\ref{tab:ERP080818} run-times
  are shown in seconds).   It would be useful to improve the run-times, perhaps
  by better parallelization.
  
  \Pbreak
  A natural question for follow-up work: are the ERP's predictive? The likely answer is that they're
  \emph{weakly} predictive. Our historical CRRA fits suggest, but do not prove, that our forward-looking
   ERP's have predictive power. Another argument for weak predictive power follows from the clear association of our ERP estimates with the volatility environment. For example, in my experience, time series (stochastic process) models with volatility-in-mean are weak predictors of future S\&P 500 returns. 
  
  The theoretical rationale for some predictive ability is a common one: risk-averse investors will discount prices to
  accommodate increased volatility. At the same time, because there is a risk-return trade-off -- higher expected returns are associated with higher volatility  -- the forecasting power is weak.

   \Pbreak
  Another natural question, asked by some early readers: what is the relation of our approach
  to `Ross Recovery' \cite{ross:2015}? This is an alternative program which, like ours, seeks to infer real-world probabilities from option prices. However, beyond that common goal, there are significant, and perhaps unbridgeable differences.   
  Ross's method seems to require, in its greatest generality, a recurrent time-homogeneous Markov process:
  see, \cite{carryu:2012}, \cite{walden:2016}, and \cite{qinlin:2016}.
  
  A restriction to recurrent processes seems highly problematic for stock price models. Prices
  are best modeled with long-run exponential growth and so are transient processes: they need not revisit their states.  Here, 
  while we don't even introduce a  dynamical stochastic process, we are certainly working in
  an unbounded continuous state space: the positive axis for $S_T$. Moreover, our transition densities are 
  (conditionally) translationally invariant in $x_t = \log S_t$; i.e., stock price level-independent. That property leads to 
   $q/p \propto \e^{-\kappa (x_t - x_0)}$. The bottom line is that, if sensible stock
  price models are transient, that leads to ``unrecoverable" parameters, such as $\kappa$ -- that need to be estimated from stock price histories.

  \Pbreak
  Finally, as mentioned at the beginning in footnote \ref{ft:COVID}, a timely application of the methods here may
  be found in \cite{lewis:2020a}, which studies the US ERP's during the COVID-19 pandemic.

  \newpage
  
  \newpage

  \clearpage
  
  \newpage

 \clearpage

  \clearpage

  \clearpage 
  
  \clearpage
  
  \clearpage

\clearpage 

  \clearpage
  
\clearpage

   \section{Appendix A -- representative agent/power utility model}

   \subsection{General theory} \label{sec:gentheory}
   Consider a discrete-time securities market with a probability measure $\PBB$, information flow $I_t$, and
   time-$t$ conditional expectations $\Et{\cdots}$. The fundamental theorem of asset pricing provides that, under
   the absence of arbitrage opportunities, there exists a positive
   stochastic discount factor (process) $m_{t,T}$ that prices every traded security via:\footnote{A careful textbook treatment is
   	found in \cite{back:2010}.}  
   
   \be P_t  =  \Et{m_{t,T} P_T}, \quad \mbox{for all} \,\, t \le T < \infty. \label{eq:optionpricing0} \eb
   Note that $m_{t,T}$ is a random variable at time $t$, whose value is fully revealed by time $T$; thus $m_{t,T} \in I_T$.
   Critically, the \emph{same} stochastic discount factor process prices every security.    
   Recalling our earlier notation, for risk-free discount bonds maturing at $T$, $P_T = 1$, and so
   \be D_{t,T} = \frac{1}{1 + R^f_{t,T}} = \e^{-r_{t,T} (T-t)} = \Et{m_{t,T}}. \label{eq:defaultfree} \eb
   Thus (\ref{eq:optionpricing0}) is equivalent to a discounted expectation under an inferred $\QBB$-measure, via
   \be P_t  =  D_{t,T} \, \frac{\Et{m_{t,T} P_T}}{\Et{m_{t,T}}} =  D_{t,T} \, \Eqt{P_T}. \label{sdfpricing} \eb
   Next, our representative agent model fleshes this out -- makes explicit $m_{t,T}$ -- and we'll show leads to the previous
   relation (\ref{eq:pqduality}) for the change of measures.

   \subsection{A representative agent with power utility} 
   We assume security prices $\{P_t\}$ are set by a representative
   agent with time-$t$ (`today') wealth $W_t$, investment horizon $T > t$, who makes portfolio choices to maximize an expected 
   power utility function of return:
   
   \be  \bar{U}(W_t) = \e^{-\rho (T-t)} \, \Et{ \frac{1}{1 - \kappa} \left( \frac{W_T}{W_t} \right)^{1-\kappa}}. \eb    
   Here $\rho \ge 0$ is a time-impatience rate and $\kappa \ge 0$ is the Arrow-Pratt CRRA.  When $\kappa > 0$, the agent is strictly risk-averse, and when $\kappa = 0$, risk-neutral. When $\kappa = 1$ the term inside the bracket is interpreted as $\log (W_T/W_t)$. 
   At time-$t$, the agent's investment universe contains three traded securities:
   
   \begin{itemize}
   	\item A risk-free bond, here a contract between individuals, with deterministic (non-random) 
   	price $D_{t,T}=\e^{-r_{t,T} (T-t)}$. 
   	\item A broad-based positive (equity) market index $\bar{S}_t$, with future random price $\bar{S}_T \in (0,\infty)$. This is a \emph{total-return} index, which includes reinvested dividends.  
   	\item A generic contingent claim, also a contract between individuals, with today's price $V_{t,T}$,
   	maturity $T$, and buyer payoff $V_T = w(\log \bar{S}_T)$ with generic payoff function $w(\cdot)$.
   \end{itemize}
    Implicitly, the equity index is the sole  source of uncertainty. 
   Since the risk-free bond, as described, is a type of `generic contingent claim', it's  
   redundant in the above list. But there is no harm singling it out. 
   The agent optimally allocates wealth fractions: $x$ to the discount bond,
   $y$ to the generic contingent claim, and $1-x-y$ to stocks.  The agent's optimization problem is:
   
   \be \max_{x,y} \bar{U}(x,y; S_t) = \max_{x,y} 
   \Et{ \frac{1}{1 - \kappa}  \left( (1-x-y) \, \frac{\bar{S}_T}{\bar{S}_t} + \frac{x}{D_{t,T}} + y \frac{V_T}{V_{t,T}} \right)^{1-\kappa}} \e^{-\rho (T-t)}.  \eb
   Next, we assume a stable, market-clearing equilibrium exists. By that, we mean the following two conditions hold.
   First, $f(\epsilon) \equiv \bar{U}(1-\epsilon,0;S_t)$ is a twice-differentiable, concave function of $\epsilon$
   for all $\epsilon \in [0,1]$, strictly increasing for $\epsilon \in [0,1)$. Second, the agent's optimal solution for
    $\bar{U}(x,y; S_t)$ lies at $x^* = y^* = 0$. In other words, the equity index must absorb 100\% of the
   representative agent's wealth $W_t$, and any other allocation is inferior. The other securities (the discount bond and the derivative) do exist and are traded in this market. But as contracts between individuals (like puts, calls, and futures), they are in zero net supply.  
   
   The first condition implies $f'(\epsilon=0) > 0  \Rightarrow \Et{\bar{S}_T/\bar{S}_t} >  1 + R^f_{t,T}$; i.e., the equity risk premium, $\mbox{ERP}_{t,T} \equiv \Et{R^e_{t,T} - R^f_{t,T}}$, must be positive. Concavity implies $f''(\epsilon) < 0$, where
   \be f''(\epsilon) = -\kappa \, \Et{ \left( \epsilon \, \frac{\bar{S}_T}{\bar{S}_t} + \frac{1 - \epsilon}{D_{t,T}} \right)^{-1-\kappa}
   	                    \left( \frac{\bar{S}_T}{\bar{S}_t} - \frac{1}{D_{t,T}} \right)^2},
                       \quad (0 \le \epsilon \le 1).  \label{eq:ftwo} \eb
   By assumption, the expectation $\Et{\cdots}$ in (\ref{eq:ftwo}) exists, and it's seen to be manifestly positive; our stability conditions then require that $\kappa > 0$:  the agent is strictly risk-averse.                   
   
   The second condition implies the two first order conditions hold:
   
   \[  (i) \,\, \left. \frac{\partial}{\partial x}  \bar{U} \right|_{x=y=0} = 0 \quad
   \mbox{and} \quad (ii) \,\, \left. \frac{\partial}{\partial y}  \bar{U} \right|_{x=y=0} = 0,   \] 
   which yields:   
   \be D_{t,T} = \frac{\Et{ \left(\frac{\bar{S}_T}{\bar{S}_t} \right)^{-\kappa}}}
   {\Et{ \left(\frac{\bar{S}_T}{\bar{S}_t} \right)^{1-\kappa}}} \quad \mbox{and} \quad                        
   V_{t,T} = \frac{\Et{ \left(\frac{\bar{S}_T}{S_t} \right)^{-\kappa} V_T}}
   {\Et{ \left(\frac{\bar{S}_T}{\bar{S}_t} \right)^{1-\kappa}}}. \label{eq:RNstd} \eb     
   Substituting the denominator for $V_{t,T}$ with the same denominator from $D_{t,T}$ yields
   
   \begin{empheq}{align}
   V_{t,T} =  D_{t,T} \, \Ethree{t}{Q}{V_T}, \quad \mbox{where} \quad
   \Ethree{t}{Q}{V_T} \equiv \frac{\Et{ \left(\frac{\bar{S}_T}{\bar{S}_t} \right)^{-\kappa} V_T}}
   {\Et{ \left(\frac{\bar{S}_T}{\bar{S}_t} \right)^{-\kappa}}}. \label{eq:RNpricing}
   \end{empheq}    
   Since the contingent claim payoff is an arbitrary function of $\bar{S}_T$, (\ref{eq:RNpricing}) is a version of (\ref{sdfpricing}). Indeed, we must have  $m_{t,T} = c_{t,T} \, (\bar{S}_T/\bar{S}_t)^{-\kappa}$, for some non-random $c_{t,T} \in I_t$.
   Since $D_{t,T} = \Et{m_{t,T}}$, comparing with the first equation in (\ref{eq:RNstd}) says 
   $c_{t,T} = 1/\Et{ (\bar{S}_T/\bar{S}_t)^{1-\kappa}}$. Summarizing, under our market-clearing equilibrium
   model with power utility,

   \begin{empheq}[box=\fbox]{align}
   m_{t,T} = c_{t,T} \left(\frac{\bar{S}_T}{\bar{S}_t} \right)^{-\kappa} = c_{t,T} \, \, \e^{-\kappa \bar{X}_T},
     \,\, \mbox{where} \,\,  c_{t,T} =  \{ \Et{ (\bar{S}_T/\bar{S}_t)^{1-\kappa}} \}^{-1}. \label{eq:sdf}
    \end{empheq}
    One can also check that, as required by (\ref{eq:optionpricing0}), we have $\bar{S}_t = \Et{m_{t,T} \bar{S}_T}$. 
    
    \Pbreak   
    Recall our previous notations: $\bar{X}_T = \log (\bar{S}_T/\bar{S}_t)$. Also, recall  
    $p_{\bar{X}_T}(x)$ and $q_{\bar{X}_T}(x)$ denote the corresponding $\PBB/\QBB$ total-return pdf's,   
     and $V_T = w(\log \bar{S}_T)$. Using those notations, the second equation in (\ref{eq:RNpricing}) may be written:
    
    \be \Ethree{t}{Q}{ w(\log \bar{S}_T)} =  \frac{ \int \e^{-\kappa x} w(x + \log \bar{S}_t) \, p_{\bar{X}_T}(x) \, dx}
    {\int \e^{-\kappa x} \, p_{\bar{X}_T}(x) \, dx}. \label{eq:RNpricing2}  \eb
    Now specialize to the particular payoff $w(\log \bar{S}_T) = \delta(\log (\bar{S}_T/\bar{S}_t) - y)$, where $y$ is an
    arbitrary real number, and $\delta(\cdot)$ is the Dirac delta. In other words,
    $w(x) = \delta(x - \log \bar{S}_t - y)$. With that payoff, (\ref{eq:RNpricing2}) reads    
     \be q_{\bar{X}_T}(y) =  \frac{ \e^{-\kappa y} \, p_{\bar{X}_T}(y)}
    {\int \e^{-\kappa x} \, p_{\bar{X}_T}(x) \, dx}. \label{eq:RNpricing3}  \eb 
    In the reverse direction, (\ref{eq:RNpricing3}) implies $p_{\bar{X}_T}(y) =  C \, \e^{\kappa y} \, q_{\bar{X}_T}(y)$,
    where $C$ is a constant independent of $y$. Since $p_{\bar{X}_T}(y)$ is a probability density, $C$ is determined and so
     \be p_{\bar{X}_T}(y) =  \frac{ \e^{\kappa y} \, q_{\bar{X}_T}(y)}
    {\int \e^{\kappa x} \, q_{\bar{X}_T}(x) \, dx}. \label{eq:RNpricing4}  \eb       
    Equations (\ref{eq:RNpricing3})-(\ref{eq:RNpricing4}) are  the density transformations used in the main
    body at (\ref{eq:pqduality}) or given formally at (\ref{pqdualformal}). They were asserted to follow from our representative agent model, and now we've shown that they do.

   \newpage

   \begin{table}[t] 
  	\caption{{\bf{ERP estimated term structure: SPX options on Feb 7, 2018.}}}
  	\begin{center}
  		\begin{tabular}{llcccclccc} 			
  			\toprule 
  			File/ & Expiration     & Put-   &     &    &            & Bid-Ask     & Convergence  & ERP     &Run \\
  		   Root & Date (days-to-go) & BidMin & N   & PG & $N_{opts}$  & OutStats & (steps) & $(\kappa=3)$ &time \\		
  			\cmidrule(r){1-1}   \cmidrule(r){2-2} \cmidrule(r){3-3} \cmidrule(r){4-4}
  			\cmidrule(r){5-5}  \cmidrule(r){6-6}   \cmidrule(r){7-7}  \cmidrule(r){8-8} \cmidrule(r){9-9} \cmidrule(r){10-10}
      1 SPXW &Feb 9, 2018 (2)    & 0.05   & 5   & 5  & 139        & A(1,0.05) & No(250)     & 26.34 & 1055 \\
      2 SPXW &Feb 12, 2018 (5)   & 0.05   & 5   & 5  & 127        & G(0,0)    & Yes(133)     & 16.21 & 484 \\
      3 SPXW &Feb 14, 2018 (7)   & 0.05   & 5   & 5  & 142        & A(1,0.01) & Yes(115)     & 16.16 & 538 \\
      4 SPX  &Feb 16, 2018 (9)   & 0.05   & 5   & 5  & 239        & A(1,0.08) & Yes(174)     & 15.73 & 1076 \\
      5 SPXW &Feb 16, 2018 (9)   & 0.05   & 5   & 5  & 228        & A(1,0.03) & No(250)     & 15.93 & 1554 \\
      6 SPXW &Feb 20, 2018 (13)  & 0.05   & 5   & 5  & 147        & A(2,0.03) & Yes(75)     &  12.26 & 312 \\
      7 SPXW &Feb 21, 2018 (14)  & 0.05   & 5   & 5  & 148        & G(0,0)    & Yes(66)     &  12.43 & 263 \\
      8 SPXW &Feb 23, 2018 (16)  & 0.05   & 5   & 5  & 185        & G(0,0)    & Yes(118)     & 12.78 & 719 \\
      9 SPXW &Feb 26, 2018 (19)  & 0.05   & 4   & 5  & 145        & G(0,0)    & Yes(198)     &  11.34 & 412 \\
      10 SPXW&Feb 28, 2018 (21)  & 0.40   & 4   & 5  & 175        & A(1,0.06) & Yes(102)     & 11.80 & 426 \\
      11 SPXW&Mar 2, 2018 (23)   & 0.05   & 4   & 5  & 199        & G(0,0)   & No(250)      & 12.14 & 985 \\
      12 SPXW&Mar 5, 2018 (26)   & 0.05   & 4   & 5  & 128        & A(1,0.11 & Yes(65)     &   10.93 & 156 \\
      13 SPXW&Mar 7, 2018 (28)   & 0.05   & 4   & 5  & 131        & G(0,0)   & Yes(44)     &  10.91 &  135 \\
      14 SPXW&Mar 9, 2018 (30)   & 0.05   & 4   & 5  & 176        & A(1,0.02) & Yes(79)     & 11.54 & 305 \\
      15 SPXW&Mar 12, 2018 (33)  & 0.05   & 4   & 5  & 127        & G(0,0)    & Yes(35)     &  10.21 & 86 \\
      16 SPXW&Mar 14, 2018 (35)  & 0.05   & 4   & 5  & 104        & G(0,0)    & Yes(103)     &  10.20 & 135 \\
      17 SPX&Mar 16, 2018 (37)   & 0.05   & 4   & 5  & 259        & G(0,0)    & Yes(77)     &  10.43 & 309 \\
      18 SPXW&Mar 16, 2018 (37)  & 0.05   & 4   & 5  & 259        & A(1,0.01) & Yes(217)     &  10.39 & 878 \\
      19 SPXW&Mar 23, 2018 (44)  & 0.05   & 4   & 5  & 159        & G(0,0)    & Yes(42)     &  10.01 & 167 \\
      20 SPXW&Mar 29, 2018 (50)  & 0.05   & 4   & 5  & 198        & G(0,0)    & Yes(59)     &  9.55 & 236 \\
      21 SPXW&Apr 6, 2018 (58)   & 0.05   & 4   & 5  & 143        & G(0,0)    & Yes(90)     &  8.93 & 333 \\
      22 SPX&Apr 20, 2018 (72)   & 0.05   & 4   & 5  & 244        & A(1,0.08) & Yes(64)     &  8.72 & 269 \\
      23 SPXW&Apr 20, 2018 (72)  & 0.05   & 5   & 5  & 246        & A(2,0.01) & Yes(156)     & 8.64 & 1046 \\
      24 SPXW&Apr 30, 2018 (82)  & 0.05   & 4   & 5  & 106        & G(0,0)    & Yes(137)     & 8.39 & 177 \\
      25 SPX&May 18, 2018 (100)  & 0.05   & 4   & 5  & 168        & G(0,0)    & Yes(35)     &  8.52 & 126 \\
      26 SPXW&May 18, 2018 (100) & 0.05   & 4   & 5  & 167        & G(0,0)    & Yes(44)     &  8.50 & 163 \\
      27 SPXW&May 31, 2018 (113) & 0.05   & 4   & 5  & 60         & G(0,0)    & Yes(73)     &  8.25 & 86 \\
      28 SPX&Jun 15, 2018 (128)  & 0.05   & 4   & 5  & 93         & A(1,0.04) & No(250)     &  8.23 & 306 \\
      29 SPXW&Jun 15, 2018 (128) & 0.05   & 4   & 5  & 95         & A(2,0.05) & Yes(95)     &  8.17 & 115 \\
      30 SPXW&Jun 29, 2018 (142) & 0.05   & 4   & 5  & 167        & G(0,0)    & Yes(29)     &  8.10 & 107 \\
      31 SPXW&Jul 31, 2018 (174) & 0.05   & 4   & 5  & 87         & G(0,0)    & Yes(34)     &  7.79 & 43 \\
      32 SPX&Sep 21, 2018 (226)  & 0.05   & 4   & 5  & 86         & G(0,0)    & Yes(109)     & 7.56 & 132 \\
      33 SPXW&Sep 21, 2018 (226) & 0.05   & 4   & 5  & 87         & G(0,0)    & Yes(46)     &  7.53 & 55 \\
      34 SPXW&Sep 28, 2018 (233) & 0.05   & 4   & 5  & 63         & G(0,0)    & Yes(90)     &  7.52 & 109 \\
      35 SPX&Dec 21, 2018 (317)  & 0.05   & 4   & 5  & 105        & A(1,0.05) & No(250)     &  7.37 & 324 \\
      36 SPXW&Dec 21, 2108 (317) & 0.05   & 4   & 5  & 98         & G(0,0)    & Yes(63)     &  7.36 & 76 \\
      37 SPXW&Dec 31, 2018 (327) & 0.05   & 4   & 5  & 62         & G(0,0)    & Yes(44)     &  7.47 & 53 \\
      38 SPX&Jan 18, 2019 (345)  & 0.05   & 4   & 5  & 89         & G(0,0)    & Yes(39)     &  7.25 & 47 \\
      39 SPX&Mar 15, 2019 (401)  & 0.05   & 4   & 5  & 85         & G(0,0)    & Yes(147)    &  7.29 & 176 \\
      40 SPX&Jun 21, 2019 (499)  & 0.05   & 4   & 5  & 92         & G(0,0)    & Yes(60)     &  7.28 & 73 \\
      41 SPX&Dec 20, 2019 (681)  & 0.30   & 4   & 5  & 93         & G(0,0)    & No(250)     &  7.23 & 309 \\
      42 SPX&Dec 18, 2020 (1045) & 0.30   & 4   & 5  & 96         & G(0,0)    & No(250)     &  7.67 & 301 \\
  			\bottomrule 
  		\end{tabular}
  	\end{center}
  	\label{tab:ERP020718}
  \end{table}

 \newpage 
 
  \begin{table}[t] 
 	\caption{{\bf{ERP estimated term structure: SPX options on Aug 8, 2018.}}}
 	\begin{center}
 		\begin{tabular}{llcccclccc} 			
 			\toprule 
 		File/ &	Expiration     & Put-   &     &    &            & Bid-Ask      & Convergence  & ERP &Run \\
 		Root &	Date (days-to-go) & BidMin & N   & PG & $N_{opts}$  & OutStats & (steps) & $(\kappa=3)$ &time \\		
 			\cmidrule(r){1-1}   \cmidrule(r){2-2} \cmidrule(r){3-3} \cmidrule(r){4-4}
 			\cmidrule(r){5-5}  \cmidrule(r){6-6}   \cmidrule(r){7-7}  \cmidrule(r){8-8} \cmidrule(r){9-9} \cmidrule(r){10-10}
 	1 SPXW&Aug 10, 2018 (2) & 0.10   & 5   & 5  & 31         & G(0,0)    & Yes(63)      & 2.02 & 120 \\
 	2 SPXW&Aug 13, 2018 (5) & 0.10   & 5   & 5  & 51         & G(0,0)    & Yes(62)     & 1.45 & 119 \\
 	3 SPXW&Aug 15, 2018 (7) & 0.10   & 5   & 5  & 87         & A(2,0.01) & Yes(76)     &  1.87 & 183 \\
 	4 SPX&Aug 17, 2018 (9)  & 0.10   & 5   & 5  & 112        & A(2,0.01) & Yes(67)     &  2.06 & 361 \\
 	5 SPXW&Aug 17, 2018 (9) & 0.10   & 5   & 5  & 119        & A(2,0.02) & Yes(73)     &  2.18 & 463 \\
 	6 SPXW&Aug 20, 2018 (12) & 0.10   & 5   & 5  & 117        & G(0,0)    & Yes(57)     &  1.99 & 356 \\
 	7 SPXW&Aug 22, 2018 (14) & 0.10   & 5   & 5  & 122        & G(0,0)    & Yes(68)     &  2.22 & 439 \\
 	8 SPXW&Aug 24, 2018 (16) & 0.10   & 5   & 5  & 162        & G(0,0)    & No(250)     &  2.43 & 1623 \\
 	9 SPXW&Aug 27, 2018 (19) & 0.10   & 5   & 5  & 121        & G(1,0.00)   & No(250)     &  2.30 & 1506 \\
 	10 SPXW&Aug 29, 2018 (21) & 0.10   & 5   & 5  & 106        & A(1,0.01) & No(250)     &  2.48 & 1316 \\
 	11 SPXW&Aug 31, 2018 (23) & 0.10   & 5   & 5  & 164        & G(1,0.00)   & No(250)    &  2.67 & 1589 \\
 	12 SPXW&Sep 4, 2018 (27) & 0.10   & 5   & 5  & 109        & A(2,0.04) & No(250)     & 2.49 & 1321 \\
 	13 SPXW&Sep 5, 2018 (28) & 0.10   & 5   & 5  & 109        & A(2,0.04) & No(250)     & 2.53 & 1288 \\
 	14 SPXW&Sep 7, 2018 (30) & 0.20   & 5   & 5  & 164        & A(1,0.02) & No(250)     & 2.70 & 1714 \\
 	15 SPXW&Sep 10, 2018 (33) & 0.10   & 5   & 5  & 111        & A(2,0.02) & No(250)     & 2.62 & 1307 \\
 	16 SPXW&Sep 12, 2018 (35) & 0.10   & 5   & 5  & 105        & G(0,0)    & No(250)     & 2.75 & 1279 \\
 	17 SPXW&Sep 14, 2018 (37) & 0.10   & 5   & 5  & 166        & A(2,0.03) & No(250)     & 2.87 & 1612 \\
 	18 SPX&Sep 21, 2018 (44) & 0.20   & 5   & 5  & 236        & A(1,0.01) & No(250)     & 3.07 & 1631 \\
 	19 SPXW&Sep 21, 2018 (44) & 0.20   & 5   & 5  & 236        & A(1,0.01  & No(250)     & 3.11 & 1601 \\
 	20 SPXW&Sep 28, 2018 (51) & 0.10   & 5   & 5  & 205        & A(1,0.03) & No(250)     & 3.33 & 1594 \\
 	21 SPXW&Oct 5, 2018 (58) & 0.10   & 5   & 5  & 151        & G(0,0)    & Yes(187)    & 3.43 & 1210 \\
 	22 SPXW&Oct 12, 2018 (65) & 0.10   & 5   & 5  & 151        & G(0,0)    & Yes(224)    & 3.52 & 1405 \\
 	23 SPX&Oct 19, 2018 (72) & 0.10   & 5   & 5  & 248        & A(1,0.03) & No(250)     & 3.60 & 1598 \\
 	24 SPXW&Oct 19, 2018 (72) & 0.10   & 5   & 5  & 247        & G(0,0)    & No(250)     & 3.63 & 1591 \\
 	25 SPXW&Oct 31, 2018 (84) & 0.10   & 5   & 5  & 167        & A(2,0.01) & Yes(121)    & 3.80 & 756 \\
 	26 SPX&Nov 16, 2018 (100) & 0.10   & 5   & 5  & 238        & A(1,0.05) & No(250)     & 4.11 & 1572 \\
 	27 SPXW&Nov 16, 2018 (100) & 0.10   & 5   & 5  & 238        & A(1,0.01) & No(250)     & 4.15 & 1589 \\
 	28 SPXW&Nov 30, 2108 (114) & 0.10   & 5   & 5  & 100        & G(1,0.00)   & Yes(108)    & 4.29 & 533 \\
 	29 SPX&Dec 21, 2018 (135) & 0.10   & 5   & 5  & 95         & A(2,0.06) & Yes(223)    & 4.44 & 571 \\
 	30 SPXW&Dec 21, 2018 (135) & 0.20   & 5   & 5  & 96         & G(0,0)    & No(250)     & 4.45 & 676 \\
 	31 SPXW&Dec 31, 2018 (145) & 0.10   & 5   & 5  & 189        & G(0,0)    & Yes(90)     & 4.42 & 564 \\
 	32 SPX&Jan 18, 2019 (163) & 0.10   & 5   & 5  & 91         & G(1,0.00)   & No(250)     & 4.54 & 663 \\
 	33 SPXW&Jan 31, 2019 (176)& 0.10   & 5   & 5  & 92         & G(0,0)    & Yes(72)     & 4.70 & 180 \\
 	34 SPX&Mar 15, 2019 (219) & 0.10   & 5   & 5  & 89         & G(0,0)    & Yes(147)    & 4.81 & 445 \\
 	35 SPXW&Mar 29, 2109 (233) & 0.10   & 5   & 5  & 87         & G(0,0)    & Yes(95)     & 4.90 & 234 \\
 	36 SPX&Jun 21, 2019 (317) & 0.10   & 5   & 5  & 94         & G(0,0)    & Yes(44)     & 5.20 & 126 \\
 	37 SPXW&Jun 21, 2019 (317) & 0.10   & 5   & 5  & 94         & G(0,0)    & Yes(32)     & 5.21 & 80 \\
 	38 SPXW&Jun 28, 2019 (324) & 0.10   & 5   & 5  & 86         & G(0,0)    & Yes(122)    & 5.28 & 301 \\
 	39 SPX&Sep 20, 2019 (408)  & 0.10   & 5   & 5  & 89         & G(0,0)    & Yes(46)     & 5.50 & 125 \\
 	40 SPX&Dec 20, 2019 (499) & 0.10   & 5   & 5  & 104        & G(0,0)    & Yes(80)     & 6.02 & 493 \\
 	41 SPX&Jun 19, 2019 (681) & 0.10   & 5   & 5  & 89         & G(0,0)    & Yes(113)    & 5.92 & 353 \\
 	42 SPX&Dec 18, 2020 (863) & 0.20   & 5   & 5  & 100        & G(0,0)    & Yes(153)    & 7.09 & 911 \\
 			\bottomrule 
 		\end{tabular}
 	\end{center}
 	\label{tab:ERP080818}
 \end{table}

\newpage

     \begin{table}[h] 
     	\caption{{\bf{Feb 7, 2018: Calendar arbitrage adjustments detail.}} \newline
     		After(+): Market prices for PM options (asterisks) are adjusted upward. \newline
     		After(-):\quad Market prices for AM options (asterisks) are adjusted downward.}              
     	\begin{center}
     		\begin{tabular}{llcccclccc}
     			I. Before   \\			 			
     			\midrule 
     			File/ & Expiration     & Put-   &     &    &            & Bid-Ask     & Convergence  & ERP     &Run \\
     			Root & Date (days-to-go) & BidMin & N   & PG & $N_{opts}$  & OutStats & (steps) & $(\kappa=3)$ &time \\		
     			\cmidrule(r){1-1}   \cmidrule(r){2-2} \cmidrule(r){3-3} \cmidrule(r){4-4}
     			\cmidrule(r){5-5}  \cmidrule(r){6-6}   \cmidrule(r){7-7}  \cmidrule(r){8-8} \cmidrule(r){9-9} \cmidrule(r){10-10}
     			17 SPX&Mar 16, 2018 (37)   & 0.05   & 4   & 5  & 259        & G(0,0)    & Yes(77)     &  10.43 & 309 \\
     			18 SPXW&Mar 16, 2018 (37)  & 0.05   & 4   & 5  & 259        & A(1,0.01) & Yes(217)     &  10.39 & 878 \\
     			35 SPX&Dec 21, 2018 (317)  & 0.05   & 4   & 5  & 105        & A(1,0.05) & No(250)     &  7.37 & 324 \\
     			36 SPXW&Dec 21, 2108 (317) & 0.05   & 4   & 5  & 98         & G(0,0)    & Yes(63)     &  7.36 & 76  \\         		
     			\midrule 
     			\\
     			\\
     			II. After(+) \\			
     			\midrule 
     			File/ & Expiration     & Put-   &     &    &            & Bid-Ask     & Convergence  & ERP     &Run \\
     			Root & Date (days-to-go) & BidMin & N   & PG & $N_{opts}$  & OutStats & (steps) & $(\kappa=3)$ &time \\		
     			\cmidrule(r){1-1}   \cmidrule(r){2-2} \cmidrule(r){3-3} \cmidrule(r){4-4}
     			\cmidrule(r){5-5}  \cmidrule(r){6-6}   \cmidrule(r){7-7}  \cmidrule(r){8-8} \cmidrule(r){9-9} \cmidrule(r){10-10}
     			17 SPX&Mar 16, 2018 (37)   & 0.05   & 4   & 5  & 259        & G(0,0)    & Yes(59)     &  10.42 & 251 \\
     			18 SPXW$^*$&Mar 16, 2018 (37)  & 0.05   & 4   & 5  & 259        & A(1,0.01) & Yes(52)     &  10.58 & 216 \\
     			35 SPX&Dec 21, 2018 (317)  & 0.15   & 4   & 5  & 97         & G(0,0)    & Yes(71)     &  7.35  & 88 \\
     			36 SPXW$^*$&Dec 21, 2108 (317) & 0.15   & 4   & 5  & 97         & G(0,0)    & Yes(85)     &  7.39 & 105 \\         		
     			\midrule 
     			\\
     			\\
     			III. After(-) \\			
     			\midrule 
     			File/ & Expiration     & Put-   &     &    &            & Bid-Ask     & Convergence  & ERP     &Run \\
     			Root & Date (days-to-go) & BidMin & N   & PG & $N_{opts}$  & OutStats & (steps) & $(\kappa=3)$ &time \\		
     			\cmidrule(r){1-1}   \cmidrule(r){2-2} \cmidrule(r){3-3} \cmidrule(r){4-4}
     			\cmidrule(r){5-5}  \cmidrule(r){6-6}   \cmidrule(r){7-7}  \cmidrule(r){8-8} \cmidrule(r){9-9} \cmidrule(r){10-10}
     			17 SPX$^*$&Mar 16, 2018 (37)   & 0.05   & 4   & 5  & 259        & G(0,0)    & Yes(90)     &  10.20 & 350 \\
     			18 SPXW&Mar 16, 2018 (37)  & 0.05   & 4   & 5  & 259        & A(1,0.01) & Yes(217)     &  10.40 & 854 \\
     			35 SPX$^*$&Dec 21, 2018 (317)  & 0.15   & 4   & 5  & 97         & G(0,0)    & Yes(73)     &  7.32  & 75 \\
     			36 SPXW&Dec 21, 2108 (317) & 0.15   & 4   & 5  & 97         & G(0,0)    & Yes(85)    &  7.36  & 88 \\         		
     			\midrule 
     		\end{tabular}
     	\end{center}
     	\label{tab:BeforeAfterCalArbAdj}
     \end{table}
     
     \clearpage

     \begin{table}[t] 
     	\caption{{\bf{ERP sensitivity to cost-of-carry method: SPX options on Feb 7, 2018.}} \newline
     	    Rates $(r,\delta,\mbox{ERP})$ shown as \%/year. Forward = $S_0 \exp((r-\delta) T)$, where $S_0 = 2706.48$.}
          	\begin{center}
     		\begin{tabular}{lcccccccccc} 			
     			\toprule 
    	File/ & $T$   &  \multicolumn{4}{c}{VIX white paper method}  & \multicolumn{4}{c}{PC-parity regression method} &Realized  \\
     				\cmidrule(r){3-6}   \cmidrule(r){7-10} 
     			Root & (Years)        & $r$ & $\delta$   & Forward & ERP   & $r$  & $\delta$  & Forward & ERP &$\delta$ \\		
     			\cmidrule(r){1-1}   \cmidrule(r){2-2} \cmidrule(r){3-3} \cmidrule(r){4-4}
 \cmidrule(r){5-5}  \cmidrule(r){6-6}   \cmidrule(r){7-7}  \cmidrule(r){8-8} \cmidrule(r){9-9} \cmidrule(r){10-10} \cmidrule(r){11-11}
   		1 SPXW  & 0.00551    & 1.24 & 10.77  & 2705.06  & \bf{26.34}         &5.69 & 14.75 & 2705.13  &  \bf{26.35} &8.36\\
   		2 SPXW  & 0.01373    & 1.25 & 6.02   & 2704.71  & \bf{16.21}         &2.91 &  7.03 & 2704.95  &  \bf{16.23} &3.59\\
   		3 SPXW  & 0.01921    & 1.26 & 5.82   & 2704.11  & \bf{16.16}         &5.44 &  9.55 & 2704.35  &  \bf{16.25} &4.29\\
       	4 SPX   & 0.02394    & 1.27 & 5.16   & 2703.96  & \bf{15.73}         &3.33 &  7.55 & 2703.73  &  \bf{15.75} &4.69\\
       	5 SPXW  & 0.02469    & 1.27 & 5.64   & 2703.56  & \bf{15.93}         &1.70 &  5.80 & 2703.74  &  \bf{15.92} &4.69\\
       	6 SPXW  & 0.03565    & 1.28 & 3.59   & 2704.26  & \bf{12.26}         &2.88 &  4.91 & 2704.53  &  \bf{12.30} &3.45\\
     	7 SPXW  & 0.03838    & 1.29 & 3.19   & 2704.51  & \bf{12.43}         &3.05 &  4.76 & 2704.70  &  \bf{12.43} &3.25\\
       	8 SPXW  & 0.04386    & 1.30 & 2.75   & 2704.76  & \bf{12.78}         &2.74 &  4.13 & 2704.84  &  \bf{12.79} &3.18\\	 
       	9 SPXW  & 0.05208    & 1.31 & 2.78   & 2704.41  & \bf{11.34}         &2.72 &  4.01 & 2704.67  &  \bf{11.41} &2.92\\	 
    	10 SPXW  & 0.05756   & 1.32 & 2.71   & 2704.31  & \bf{11.80}         &2.09 &  3.36 & 2704.50  &  \bf{11.82} &3.07\\	
     	11 SPXW  & 0.06304   & 1.32 & 2.86   & 2703.86  & \bf{12.14}         &2.48 &  3.92 & 2704.02  &  \bf{12.16} &3.20\\	
       	12 SPXW  & 0.07126   & 1.34 & 2.51   & 2704.21  & \bf{10.93}         &2.75 &  3.67 & 2704.72  &  \bf{10.94} &2.84\\			
       	13 SPXW  & 0.07674   & 1.34 & 2.36   & 2704.36  & \bf{10.91}         &2.39 &  3.37 & 2704.44  &  \bf{10.94} &2.80 \\	
   		14 SPXW  & 0.08222   & 1.35 & 2.35   & 2704.26  & \bf{11.54}         &2.33 &  3.28 & 2704.37  &  \bf{11.56} &2.97\\	
    	15 SPXW  & 0.09044   & 1.36 & 2.21   & 2704.41  & \bf{10.21}         &2.18 &  2.98 & 2704.52  &  \bf{10.24} &2.71\\	
       	16 SPXW  & 0.09592   & 1.37 & 2.36   & 2703.91  & \bf{10.20}         &2.22 &  3.14 & 2704.11  &  \bf{10.21} &2.91\\	
      	17 SPX   & 0.10066   & 1.38 & 2.38   & 2703.76  & \bf{10.43}         &2.41 &  3.27 & 2704.14  &  \bf{10.45} &2.85 \\	
       	18 SPXW  & 0.10140   & 1.38 & 2.31   & 2703.91  & \bf{10.39}         &2.35 &  3.25 & 2704.03  &  \bf{10.42} &2.85\\	
       	19 SPXW  & 0.12058   & 1.40 & 1.77   & 2705.26  & \bf{10.01}         &2.33 &  2.64 & 2705.48  &  \bf{10.03} &2.54 \\	
       	20 SPXW  & 0.13702   & 1.42 & 1.78   & 2705.16  & \bf{9.55}          &2.31 &  2.61 & 2705.38  &  \bf{9.57}  &2.39 \\	
     	21 SPXW  & 0.15893   & 1.45 & 1.54   & 2706.06  & \bf{8.93}          &2.30 &  2.36 & 2706.22  &  \bf{8.96}  &2.24\\	
     	22 SPX   & 0.19655   & 1.49 & 1.46   & 2706.61  & \bf{8.72}          &2.37 &  2.31 & 2706.80  &  \bf{8.75}  &2.15\\	
     	23 SPXW  & 0.19729   & 1.49 & 1.46   & 2706.61  & \bf{8.64}          &2.39 &  2.34 & 2706.74  &  \bf{8.67}  &2.15\\	
     	24 SPXW  & 0.22469   & 1.52 & 1.33   & 2707.60  & \bf{8.39}          &2.40 &  2.20 & 2707.70  &  \bf{8.42}  &1.94 \\	
     	25 SPX   & 0.27326   & 1.56 & 1.52   & 2706.80  & \bf{8.52}          &2.43 &  2.38 & 2706.85  &  \bf{8.56}  &2.23 \\	
    	26 SPXW  & 0.27400   & 1.56 & 1.51   & 2706.90  & \bf{8.50}          &2.46 &  2.41 & 2706.83  &  \bf{8.56}  &2.23 \\	
      	27 SPXW  & 0.30962   & 1.69 & 1.55   & 2706.84  & \bf{8.25}          &2.30 &  2.21 & 2707.24  &  \bf{8.31}  &2.19 \\	
     	28 SPX   & 0.34997   & 1.62 & 1.55   & 2707.15  & \bf{8.23}          &2.42 &  2.35 & 2707.20  &  \bf{8.28}  &2.20 \\	
       	29 SPXW  & 0.35071   & 1.62 & 1.55   & 2707.20  & \bf{8.17}          &2.28 &  2.23 & 2706.91  &  \bf{8.21}  &2.20 \\	
        30 SPXW  & 0.38907   & 1.65 & 1.45   & 2708.60  & \bf{8.10}          &2.36 &  2.15 & 2708.69  &  \bf{8.15}  &2.08 \\	
        31 SPXW  & 0.47674   & 1.70 & 1.34   & 2711.22  & \bf{7.79}          &2.36 &  1.97 & 2711.54  &  \bf{7.87}  &1.94 \\	
        32 SPX   & 0.61847   & 1.77 & 1.37   & 2713.13  & \bf{7.56}          &2.44 &  2.06 & 2712.93  &  \bf{7.62}  &2.06 \\	
        33 SPXW  & 0.61921   & 1.77 & 1.39   & 2712.88  & \bf{7.53}          &2.37 &  1.98 & 2713.05  &  \bf{7.58}  &2.06\\	
        34 SPXW  & 0.63839   & 1.78 & 1.39   & 2713.13  & \bf{7.52}          &2.36 &  1.97 & 2713.22  &  \bf{7.58}  &2.05 \\	
        35 SPX   & 0.86778   & 1.86 & 1.36   & 2718.10  & \bf{7.37}          &2.43 &  1.94 & 2718.11  &  \bf{7.46}  &2.05 \\	
        36 SPXW  & 0.86852   & 1.86 & 1.34   & 2718.76  & \bf{7.36}          &2.41 &  1.90 & 2718.68  &  \bf{7.43}  &2.05 \\	
        37 SPXW  & 0.89592   & 1.86 & 1.26   & 2721.30  & \bf{7.47}          &2.49 &  1.88 & 2721.31  &  \bf{7.54}  &2.04 \\	
        38 SPX   & 0.94449   & 1.88 & 1.27   & 2722.01  & \bf{7.25}          &2.51 &  1.90 & 2722.19  &  \bf{7.33}  &2.02 \\	
        39 SPX   & 1.09792   & 1.92 & 1.33   & 2723.99  & \bf{7.29}          &2.44 &  1.85 & 2724.18  &  \bf{7.37}  &2.07\\	
        40 SPX   & 1.36641   & 1.98 & 1.36   & 2729.63  & \bf{7.28}          &2.44 &  1.80 & 2730.22  &  \bf{7.39}  &2.07 \\	
        41 SPX   & 1.86504   & 2.10 & 1.40   & 2741.85  & \bf{7.23}          &2.44 &  1.73 & 2742.68  &  \bf{7.29}  & NA \\	
        42 SPX   & 2.86230   & 2.29 & 1.44   & 2772.82  & \bf{7.67}          &2.54 &  1.69 & 2772.77  &  \bf{7.78}  & NA \\	
     	     \bottomrule 
     		\end{tabular}
     	\end{center}
     	\label{tab:ERP020718CoC}
     \end{table}

\clearpage

 \begin{table}[t] 
	\caption{{\bf{ERP sensitivity to cost-of-carry method: SPX options on Aug 8, 2018.}} \newline
		Rates $(r,\delta,\mbox{ERP})$ shown as \%/year. Forward = $S_0 \exp((r-\delta) T)$, where $S_0 = 2861.59$.}
	\begin{center}
		\begin{tabular}{lccccccccc} 			
			\toprule 
			File/ & $T$      &  \multicolumn{4}{c}{VIX white paper method}  & \multicolumn{4}{c}{PC-parity regression method}  \\
			\cmidrule(r){3-6}   \cmidrule(r){7-10} 
			Root & (Years)        & $r$ & $\delta$   & Forward & ERP   & $r$  & $\delta$  & Forward & ERP \\		
			\cmidrule(r){1-1}   \cmidrule(r){2-2} \cmidrule(r){3-3} \cmidrule(r){4-4}
			\cmidrule(r){5-5}  \cmidrule(r){6-6}   \cmidrule(r){7-7}  \cmidrule(r){8-8} \cmidrule(r){9-9} \cmidrule(r){10-10}
			1 SPXW  & 0.00551    & 1.85 & 7.63   & 2860.68  & \bf{2.02}         &5.75 &  11.90 & 2860.62 &  \bf{2.01} \\
			2 SPXW  & 0.01373    & 1.86 & 3.67   & 2860.88  & \bf{1.45}         &1.95 &   3.95 & 2860.80  &  \bf{1.45} \\
			3 SPXW  & 0.01921    & 1.86 & 4.06   & 2860.38  & \bf{1.87}         &2.13 &   4.41 & 2860.34  &  \bf{1.87} \\
			4 SPX   & 0.02394    & 1.87 & 4.22   & 2859.98  & \bf{2.06}         &2.66 &   4.98 & 2860.00  &  \bf{2.06} \\
		      $\cdots$ \\
			39 SPX   & 1.11709   & 2.44 & 1.57   & 2889.55  & \bf{5.50}          &2.85 &  1.97 & 2889.55  &  \bf{5.54}      \\	
			40 SPX   & 1.36641   & 2.50 & 1.57   & 2898.22  & \bf{6.02}          &2.88 &  1.80 & 2898.19  &  \bf{5.83}     \\	
			41 SPX   & 1.86504   & 2.62 & 1.60   & 2916.26  & \bf{5.92}          &3.01 &  1.99 & 2916.37  &  \bf{6.09}      \\	
			42 SPX   & 2.36367   & 2.71 & 1.66   & 2933.67  & \bf{7.09}          &3.05 &  2.00 & 2934.01  &  \bf{7.00}      \\	
			\bottomrule 
		\end{tabular}
	\end{center}
	\label{tab:ERP080818CoC}
\end{table}


\end{document}